\documentclass[preprint2]{aastex631}

\usepackage{natbib}
\usepackage{amsmath}
\usepackage{color}

\shorttitle{Star formation histories for proto-cluster at $z\sim2$}
\newcommand{\zfire}{{\tt ZFIRE}}
\newcommand{\zfourge}{{\tt ZFOURGE}}
\newcommand{\illustris}{{\tt IllustrisTNG}}
\newcommand{\pros}{{\tt PROSPECTOR}}
\newcommand{\cmmnt}[1]{}

\newcommand{\logMstarMsun}{$\log[{\rm M}_{\ast}/{\rm M}_{\odot}]$}
\newcommand{\sfr}{$\rm{M}_\odot\rm{yr}^{-1}$}

\begin{document}
\title{ {\tt ZFIRE}: The Beginning of the End for Massive Galaxies at $z\sim2$ and Why Environment Matters}
\author[0000-0001-9414-6382]{Anishya Harshan}
\affiliation{School of Physics, University of New South Wales, Sydney, NSW 2052, Australia}
\affiliation{ARC Centre of Excellence for All Sky Astrophysics in 3 Dimensions (ASTRO 3D), Australia}

\author[0000-0002-8984-3666]{Anshu Gupta}
\affiliation{School of Physics, University of New South Wales, Sydney, NSW 2052, Australia}
\affiliation{International Centre for Radio Astronomy Research (ICRAR), Curtin University, Bentley WA, Australia}
\affiliation{ARC Centre of Excellence for All Sky Astrophysics in 3 Dimensions (ASTRO 3D), Australia}

\author[0000-0001-9208-2143]{Kim-Vy Tran}
\affiliation{School of Physics, University of New South Wales, Sydney, NSW 2052, Australia}
\affiliation{ARC Centre of Excellence for All Sky Astrophysics in 3 Dimensions (ASTRO 3D), Australia}

\author[0000-0002-9495-0079]{Vicente Rodriguez-Gomez}
\affiliation{Instituto  de  Radioastronomía  y  Astrofísica,  Universidad  Nacional  Autónoma  de  México,  Apdo.   Postal  72-3,  58089  Morelia, Mexico}

\author[0000-0003-1065-9274]{Annalisa Pillepich}
\affiliation{Max-Planck-Institut f$\ddot{u}$r Astronomie, K$\ddot{o}$nigstuhl 17, 69117 Heidelberg, Germany}

\author[0000-0002-2250-8687]{Leo Y. Alcorn}
\affiliation{Department of Physics and Astronomy, York University, 4700 Keele Street, Toronto, Ontario, ON MJ3 1P3, Canada}

\author[0000-0003-2804-0648]{Themiya Nanayakkara}
\affiliation{Swinburne University of Technology, Hawthorn, VIC 3122, Australia}

\author[0000-0003-1362-9302]{Glenn G. Kacprzak}
\affiliation{Swinburne University of Technology, Hawthorn, VIC 3122, Australia}
\affiliation{ARC Centre of Excellence for All Sky Astrophysics in 3 Dimensions (ASTRO 3D), Australia}

\author[0000-0002-3254-9044]{Karl Glazebrook}
\affiliation{Swinburne University of Technology, Hawthorn, VIC 3122, Australia}
\affiliation{ARC Centre of Excellence for All Sky Astrophysics in 3 Dimensions (ASTRO 3D), Australia}

\begin{abstract}

We use \zfire\ and \zfourge\ observations with the Spectral Energy Distribution (SED) fitting tool \pros\ to reconstruct the star formation histories (SFHs) of proto-cluster and field galaxies at $z\sim 2 $ and compare our results to the TNG100 run of the \illustris\ cosmological simulation suite. In the observations, we find that massive proto-cluster galaxies (\logMstarMsun$>$10.5) form $45 \pm 8 \%$ of their total stellar mass in the first $2$ Gyr of the Universe compared to $31 \pm 2 \%$ formed in the field galaxies. In both observations and simulations, massive proto-cluster galaxies have a flat/declining SFH with decreasing redshift compared to rising SFH in their field counterparts. Using \illustris, we find that massive galaxies (\logMstarMsun $\geq 10.5$)  in both environments are on average $\approx190$ Myr older than low mass galaxies (\logMstarMsun $= 9-9.5$). However, the difference in mean stellar ages of cluster and field galaxies is minimal when considering the full range in stellar mass (\logMstarMsun $\geq 9$). We explore the role of mergers in driving the SFH in \illustris\  and find that massive cluster galaxies consistently experience mergers with low gas fraction compared to other galaxies after 1 Gyr from the Big Bang. We hypothesize that the low gas fraction in the progenitors of     massive cluster galaxies is responsible for the reduced star formation.
\end{abstract}
\section{Introduction}

Star formation is one of the key processes that drives the evolution of galaxies. Understanding when, where, and how stars are formed remains one of the major goals of extragalactic astronomy. Internal processes like supernovae and Active Galactic Nuclei (AGN) feedback \citep{Efstathiou2000,Croton2006,Bower2008,Somerville2008, Fabian2012} and dynamical instabilities \citep{Kormendy2013} regulate star formation in a galaxy. Likewise, external physical processes such as galaxy-galaxy interactions, ram pressure stripping \citep{Gunn1972, Cortese2010}, strangulation \citep{Mihos1996,Moore1998,Balogh2000}, gas accretion \citep{Kacprzak2017, Tumlinson2017} and tidal interactions also leave their imprint on the star formation history of each galaxy.



Galaxies falling into the cluster potential are subjected to interactions with other galaxies and the Intra-cluster Medium (ICM). Galaxy-galaxy mergers can trigger a star formation phase in the galaxy by funneling the flow of atomic gas into the core of the galaxies \citep{Barnes1996,Springel2005, Ellison2010}. Processes like ram pressure stripping, tidal interactions and harassment deplete the star formation fuel of a galaxy. These processes cumulatively lead to a gradual decline of star formation in galaxies in the cluster environment and can be observed as the higher fraction of quenched early type galaxies in the local universe ($z<0.1$) \citep{Balogh2000,Kauffmann2004,Presotto2012a, Wetzel2013, Paccagnella2016,  Barsanti2018,Pasquali2019, Schaefer2019}.

While the effect of environment on the star formation in galaxies is well studied at $z\sim0$, the same remains ambiguous at higher redshift. Some studies find lower star formation rates in cluster members compared to isolated field galaxies at $z\sim 1$ similar to what we observe in the local universe \citep{Williams2009, Vulcani2010, Patel2011, Popesso2012,  Old2020a}. Whereas others find a reversal in this trend \citep{Elbaz2007, Peng2010,Tran2010,Elbaz2011,Muzzin2011, Wetzel2012, Allen2016}. It is still ambiguous when the environmental effects leads to significant suppression in the star formation rates of galaxies in dense regions.

The \zfire\ survey \citep{Nanayakkara2016} targets proto-cluster galaxies at $z\sim$ 2 and  1.6 selected from the \zfourge\ survey in the COSMOS and UDS fields to identify the onset of environmental effects on galaxy properties. Environmental effects on inter stellar medium (ISM) properties, such as  gas phase metallicity and electron density, appear to be not significant till $z\sim 1.5$  \citep{Alcorn2019, Kacprzak2015, Kewley2016a} similar to the results from \illustris\ \citep{Gupta2018}. However, at $z = 1.6$, there is tentative evidence of effect of environment on the star formation rates of galaxies in the proto-cluster core \citep{Tran2015} and electron density \citep{Harshan2020}.

Observations at low redshift ($z<1$) find that the massive galaxies form their stars earlier and more rapidly compared to the low mass galaxies \citep{ Cowie1996a, Brinchmann2004, Thomas2005,Treu2005,Cimatti2006, Thomas2010, Carnall2018, Webb2020}. Massive galaxies form the majority of their stars within the first 1-2 Gyr of cosmic history and start to quench as early as $z\sim3$ \citep{Straatman2014, Glazebrook2017, Forrest2019}. Similarly, galaxies in high density environments form the majority of their stars earlier compared to the field galaxies and are on average 1-2 Gyr older than the field galaxies \citep{Thomas2005}. However, at higher redshift ($z\sim 1$), the age difference between cluster and field galaxies is less significant, $\lesssim 0.5$ Gyr \citep{Webb2020}.

Cosmological simulations and semi-analytic models \citep{DeLucia2012,Furlong2015, Bahe2017, Tremmel2019, Donnari2020a, Donnari2021}  similarly show higher quenched fractions in the cluster members compared to the field sample at $z=0$ to $z = 2$. In the \illustris\  simulations, \cite{Donnari2020a, Donnari2021} find a higher quenched fraction in the low mass satellite galaxies compared to low mass centrals indicating a role of environment in quenching of low mass galaxies. On the other hand, high mass galaxies, whether they are centrals or satellites, have high quenched fractions indicating effects of both secular and environmental quenching mechanisms.



One straightforward way of studying the evolution of galaxies is to study the star formation histories (SFH) of galaxies. The reconstruction of SFHs allows us to study the stellar mass assembly and gas accretion histories of galaxies over cosmic time.  However, inferring the SFHs from observables is an extremely complex process.  Star formation histories can be reconstructed by fitting the spectral energy distributions (SED) models for different stellar populations to the observed photometry of galaxies. We have moved from a simple exponential to more complex functional forms \citep{Buat2008,Maraston2010, Papovich2011} such as lognormals \citep{Gladders2013,Abramson2015,Carnall2018} even to non parametric SFHs \citep{CidFernandas2005, Ocvirk2006, Kelson2014, Leja2017, Chauke2018, Robotham2020a}.

In this paper we will study the effect of environment on the star formation histories of galaxies in a COSMOS proto-cluster at $z=2.095$ \citep{Spitler2012,yuan2014}. We present the first measurement of SFHs in the proto-cluster environment at $z=2$. We use the SED fitting code \pros\ \citep{Leja2017,Johnson2019} in conjunction with the extensive photometric data from the \zfourge\ survey \citep{Straatman2016} and spectroscopic redshifts from the \zfire survey \citep{Tran2015, Nanayakkara2016} to reconstruct the SFHs. We study the correlation of SFHs with the stellar mass and the environment of the galaxy. We then compare our results from observations to the SFHs retrieved from the cosmological hydrodynamical simulations \illustris. 


This paper is organised as follows. In Section \ref{sec:data} we describe the data used with \pros\ to create SFHs as described in Section \ref{sec:pros}. In Section \ref{sec:illustris} we describe the SFHs from \illustris. In Sections \ref{sec:results} and \ref{sec:discussion} we state our results and discussion and in Section \ref{sec:summary} summarise the results. For this work, we assume a flat $\Lambda CDM$ cosmology with $\Omega_{M}=0.3$, $\Omega_{\Lambda}=0.7$, and $H_0=69.6\, \rm{km s}^{-1} \rm{Mpc}^{-1}$.


\section{Methodology}
\subsection{\zfire\ and \zfourge\ Surveys}
\label{sec:data}
Our observation sample is taken from the \zfourge\ - Fourstar Galaxy Evolution Survey \citep{Straatman2016}, a deep, UV to FIR, medium-band survey, completed on the FourStar instrument \citep{persson2013} on the Magellan Telescope. It reaches depth of $\sim 26 $ mag in $J_1, J_2, J_3$ and $\sim 25$ mag in $H_s, H_l , K_s$ bands. \zfourge\ spans the Cosmic  Evolution  Survey  field \citep[COSMOS]{Scoville2007} that covers the spectroscopically confirmed proto-cluster at $z_{cl}=2.09 \pm 0.00578$ \citep{Spitler2012, yuan2014}. The \zfourge\ survey reaches $80\%$ completeness till 25.5 AB magnitude in K$_s$ band, which corresponds to \logMstarMsun\ = 9 at $z=2$ \citep{Straatman2016}. The photometric redshifts and stellar mass catalogs were created with the publicly available SED fitting codes EAZY \citep{Brammer2008} and FAST \citep{Kriek2009}.  The UV-IR star formation rates were derived in \cite{Tomczak2016}.

The \zfire\ survey \citep{Tran2015, Nanayakkara2016} is the spectroscopic follow-up of \zfourge\ survey using Keck-MOSFIRE \citep{Mclean2010,Mclean2012}. The spectroscopic sample was selected based on the \zfourge\ photometric redshifts of the proto-cluster discovered at $z\sim2$ \citep{Spitler2012}. The estimated halo mass of the $z=2.09$ COSMOS proto-cluster based on the velocity dispersion measurements has virial mass in range $\rm{M}_{vir} = 10^{13.5\pm0.2} \rm{M}_\odot$. More than two emission lines: $H\alpha, H\beta, [NII]$ or $[OIII]$ observed in H and K bands are used to measure the spectroscopic redshifts.

\begin{figure}
    \noindent
    
    \includegraphics[scale = 0.345]{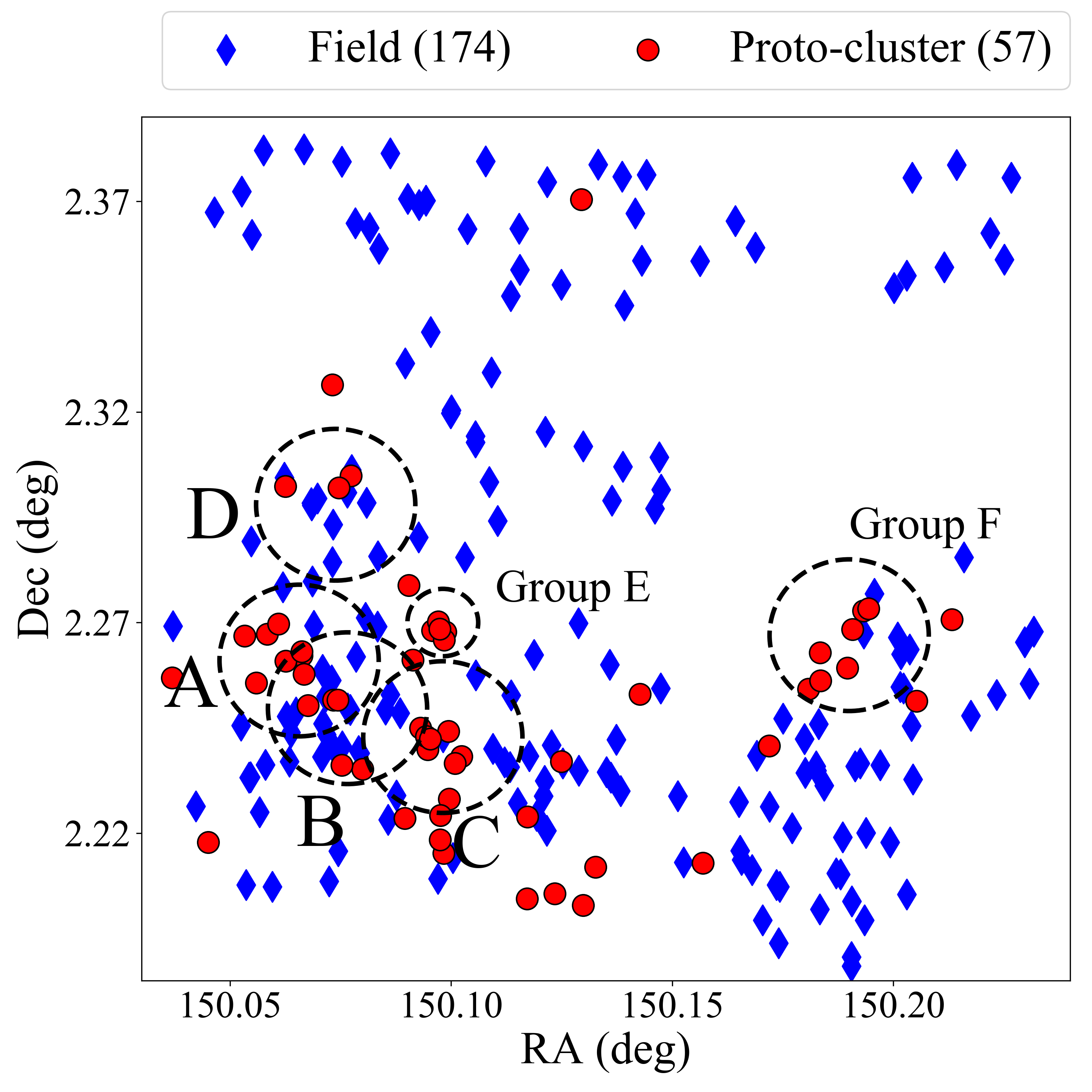}
    \caption{Spatial Distribution of spectroscopically confirmed \zfire\ proto-cluster (red circle) and field (blue diamonds) galaxies in the COSMOS field at $z\sim2$.  The dashed rings indicate the peaks in surface density as identified by \cite{yuan2014} and \cite{Spitler2012}. }
    \label{fig:radec}
\end{figure}

\begin{figure}
    \noindent
     \includegraphics[scale=0.345]{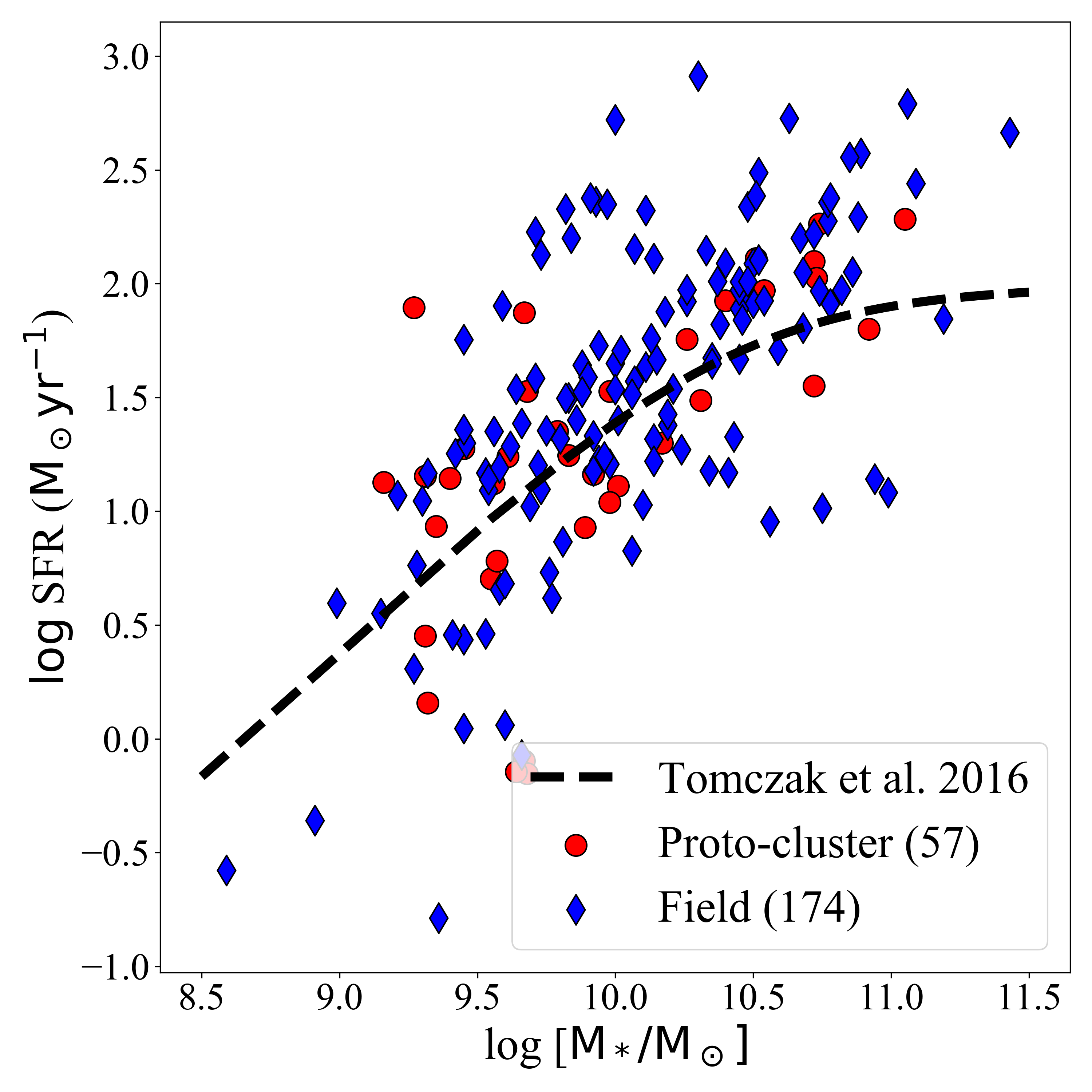}
    \caption{Stellar Mass - SFR relation of selected cluster (red circle) and field (blue diamonds) galaxies from \zfire. Stellar Mass and the UV-IR SFR has been taken from the \zfourge\ survey \citep{Straatman2016, Tomczak2016}}
    \label{fig:sfrmass}
\end{figure}

We select 57 cluster members in the COSMOS proto-cluster between redshift $2.08 < z < 2.12$ \citep{Spitler2012,yuan2014, Tran2015} and 130 field galaxies in $1.8 < z < 2.5$. Figure \ref{fig:radec} shows the spatial distribution of the spectroscopically confirmed proto-cluster (red circles) and field (blue diamonds) across the COSMOS field. This distrubution is not a depiction of true distribution of galaxies in $1.8 < z < 2.5$ in the \zfourge\ survey, but an effect of the observational strategy used in the spectroscopic follow-up for the \zfire\ survey \citep{yuan2014, Nanayakkara2016}. Figure \ref{fig:sfrmass} shows the SFR-Stellar mass relation of the selected cluster (red circles) and field (blue diamonds) galaxies. Due to the observational limit on the H$\alpha$ flux, \zfire\ observations have a SFR lower limit of $0.8$ \sfr\ \citep{yuan2014}. 


\begin{figure*}

	\includegraphics[scale=0.24]{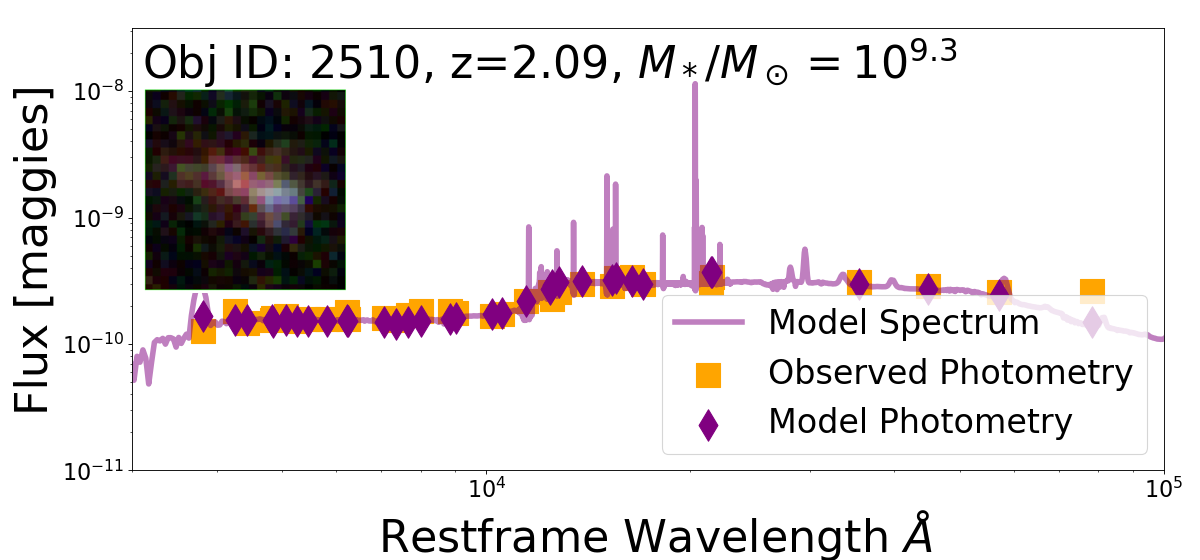}
	\includegraphics[scale=0.24]{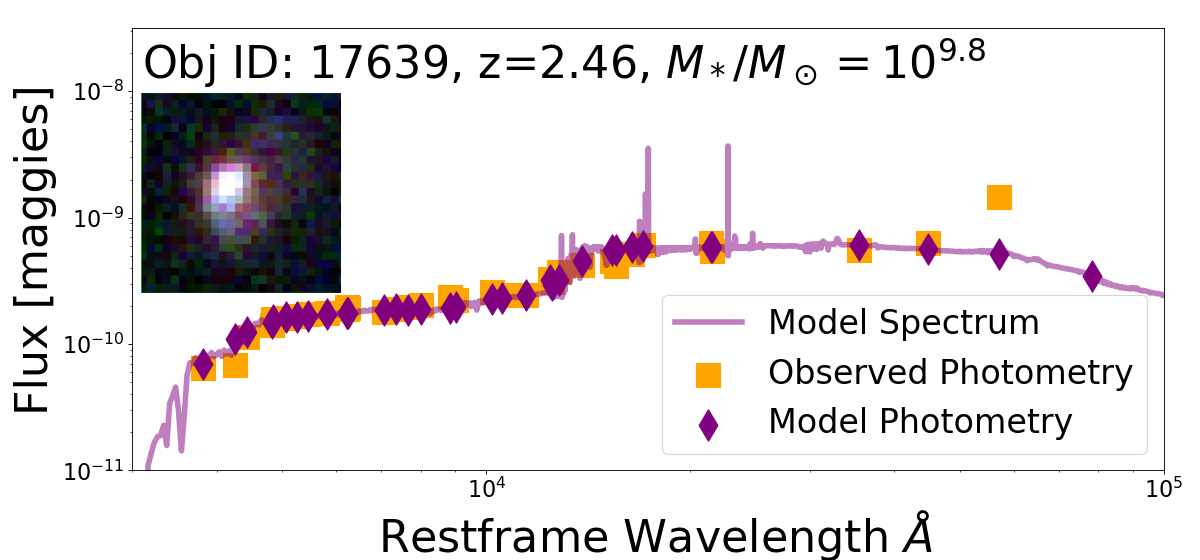}\\
	\includegraphics[scale=0.24]{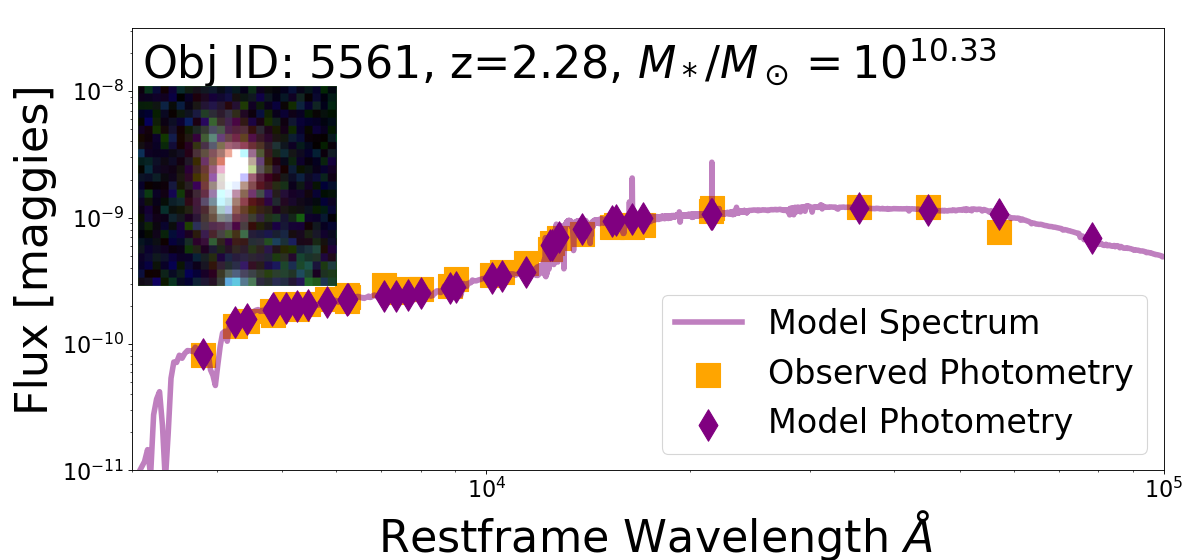}
	\includegraphics[scale=0.24]{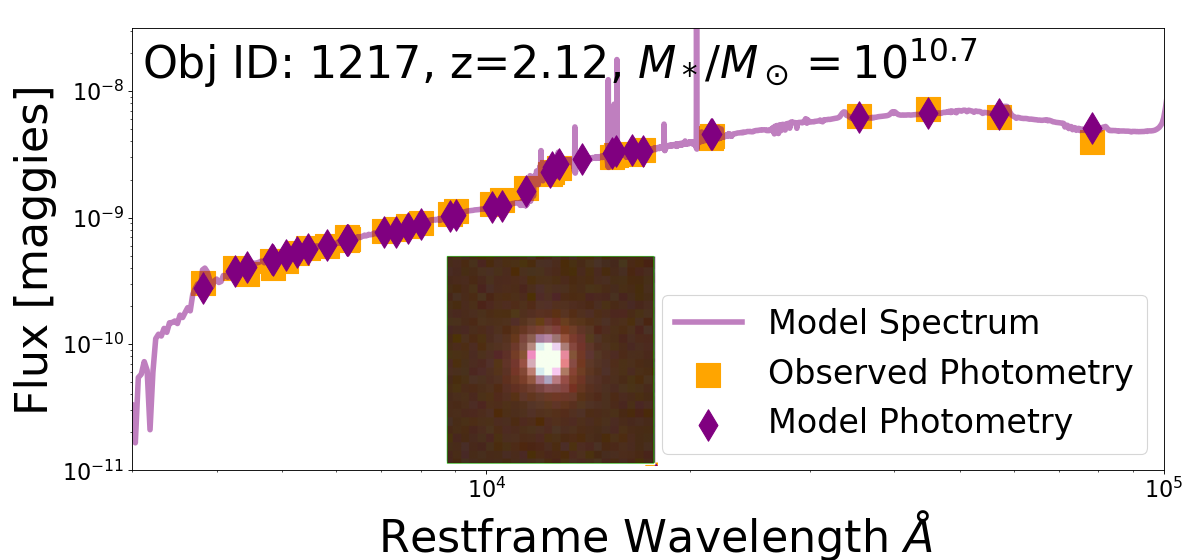}
    
    \caption{Examples of observed \zfourge\ photometry (orange squares) fitted with \pros\ to get modelled photometry (purple diamonds) and modelled SED (purple line). The stamps show 1.5$\arcsec$x1.5$\arcsec$ RGB (F160W, F814W, F606W) images of galaxies from 3D-HST survey. }
    \label{fig:SED}

\end{figure*}

\subsection{Star Formation Histories using \pros\ SED Fitting}
\label{sec:pros}
\pros\ is a SED fitting tool used to derive physical properties of galaxies using photometry and spectra. \pros\  uses the python-Flexible Stellar Population Synthesis package \citep{Conroy2009} and we use the MESA Isochrones and Stellar Tracks (MIST; \cite{Dotter2016,Paxton2015,Paxton2013,Paxton2011,Calzetti1994} and takes into account the nebular emission \citep{Byler2018}, dust attenuation and re-radiation. It uses Bayesian inference framework to derive non parametric formulation of star formation histories using  simple piece-wise constant functions. \pros\ also allows for adaptive time binning for the SFHs with varying number of bins. It fits non-parametric SFHs by calculating the fraction of stellar mass formed in a particular time bin \citep{Leja2019}. We use Calzetti dust attenuation model \citep{Calzetti1994}, Chabrier IMF \citep{Chabrier2003} and WMAP9 \citep{Hinshaw2013} cosmology throughout the analysis.

We use \pros\ on the 5 NIR medium-band photometry from the \zfourge\ survey along with 32 other UV-MIR photometric bands from the legacy data sets covering the wavelength regime of $0.4448 - 7.9158\, \mu m $ \citep{Straatman2016} for the cluster and field galaxies and spectroscopic redshifts from \zfire. We use a uniform prior across all stellar mass bin allowing us to do a comparative analysis of the SFHs. We keep 9 free parameters: stellar mass, stellar and gas-phase metallicity, dust attenuation and five independent non-parametric SFH bins. We choose the age bins to roughly match the time resolution of the \illustris\ cosmological simulation \citep{Nelson2019a} (described in Section \ref{sec:illustris}). We use the following 5 time bins : 0-200 Myr, 200-400 Myr, 400-600 Myr, 600-1000  Myr and 1000- $(t_{univ} - 1000)/2$ Myr. With the prescribed age bins, \pros\ fits for 6 SFH bins, but the additional constraint on fractional stellar mass to be summed to one results in only five independent SFH bins. Our results do not depend significantly on the choice of age bins. 

The stellar mass of galaxies calculated from \pros\ agrees reasonably with the stellar masses from FAST from the \zfourge\ survey. There is a reasonable agreement between the two with higher stellar masses of galaxies from \pros\ as seen and discussed in \cite{Leja2019}. This is speculated to be a result of different assumptions and models for SFHs in FAST and \pros.

We fit SEDs of 57 cluster galaxies in the redshift regime $2.08 \geq z \geq 2.12$ and 130 field galaxies in the redshift regime $1.8 \geq z \geq 2.5$ using the \zfire\ spectroscopic redshifts. We extract the posterior distribution of the stellar mass, five SFHs time bins and take the 16$^{\rm th}$, 50$^{\rm th}$ and 84$^{\rm th}$ percentile of the distribution from \pros. Figure \ref{fig:SED} shows the SED fits of four galaxies. The model spectrum (solid purple line) and the model photometry (purple diamonds) are well fitted to the observed photometry (orange squares).

We divide the proto-cluster and field galaxies into four stellar mass bins: $ 9 - 9.5$ \logMstarMsun, $ 9.5 - 10$  \logMstarMsun, $ 10 - 10.5$ \logMstarMsun\ and  \logMstarMsun $\geq 10.5 $. We create boostrapped samples of individual SFHs in each stellar mass bins and present the medians and errors in the median in Figure \ref{fig:sfh_pros}.

\subsection{Star Formation Histories from \illustris}
\label{sec:illustris}
\illustris\ is a suite of magneto-hydro-dynamical cosmological simulations based on the $\lambda$-CDM cosmology \citep{Pillepich2018b, Nelson2018, Springel2017,  Marinacci2017, Naiman2017}. \illustris\ extends the Illustris framework with  kinetic  black  hole  feedback,  magneto-hydrodynamics, and a revised scheme for galactic winds, among other changes \citep{Weinberger2017, Pillepich2018a}. 

In this paper we use the TNG100 box with $\rm{L}_{box} = 110.7$ cMpc, and total volume $\sim10^6 ~\rm{Mpc}^3 $, to map the star formation rate histories of the cluster and field galaxies. The data of TNG100 have been made publicly available and is described by \cite{Nelson2019a}. The TNG100 simulation has a baryonic mass resolution of $\rm{m}_b = 1.4\times 10^6 ~\rm{M}_\odot$. This resolution provides about 1000 stellar particles per galaxy for a galaxy with stellar mass \logMstarMsun $= 9$, and proportionally more stellar particles for more massive galaxies. For the entire analysis, we constrain our galaxies to have \logMstarMsun $\geq 9$ at redshift $z=2$. 

We define galaxy clusters as halos of total mass $\rm{M}_{200} \geq10^{13}\, \rm{M}_\odot$ at $z=2$ and the galaxies residing in the cluster halo except the most massive central galaxy are satellites. We identify galaxies that reside in halos of total mass $\rm{M}_{200} < 10^{13}\, \rm{M}_\odot$ at $z=2$ as field galaxies. In this analysis, galaxies are subhalos from the {\sc subfind} catalog with stellar mass \logMstarMsun $\geq 9 $ (in twice the half mass radius) at $z=2$. We select galaxies (centrals+satellites) associated with the cluster halos as cluster members whereas the field sample is made of galaxies associated with the field halos. \cite{Donnari2019} show that the main sequence of star forming galaxies in \illustris\ is lower compared to the observationally-inferred star formation main sequence at $z>\sim0.75$. Hence, to match the SFR threshold from the observations to simulations we do the following. We calculate the difference between the star formation main sequence at $z=2$ from \cite{Tomczak2016} and the observational SFR threshold of  $0.8\, \rm{M}_*\rm{yr}^{-1}$(Section \ref{sec:data}) as a function of stellar mass. We subtract the calculated difference from the median SFR-stellar mass relation of the entire sample of TNG100 and get a SFR threshold of $0.4$ \sfr. Following this selection criteria, we get a sample of 232 cluster galaxies from 24 cluster halos and 9577 field galaxies.  

We use the {\sc sublink} algorithm which tracks the merger histories of the galaxies to quantitatively follow the evolution of galaxies \citep{Rodriguez-Gomez2015}. We trace the SFRs of the cluster and field galaxies in the four defined stellar mass bins from $z=2$ to $z=6$. Figure \ref{fig:sfh_illustris} shows the distribution of the SFHs of the selected sample sets.

\section{Results}
\label{sec:results}
In this Section we show and discuss the SFHs of galaxies in a proto-cluster at $z\sim2$ derived from SED fitting of observations with \pros. We compare our results with predictions from \illustris\ and discuss the dependence of SFH on stellar mass and environment of the galaxy. We also discuss possible physical motivators to describe our findings.

\subsection{Star Formation Histories and Environment}
\label{sec:SFHvsenv}
\begin{figure*}
    \noindent
    \includegraphics[scale = 0.32]{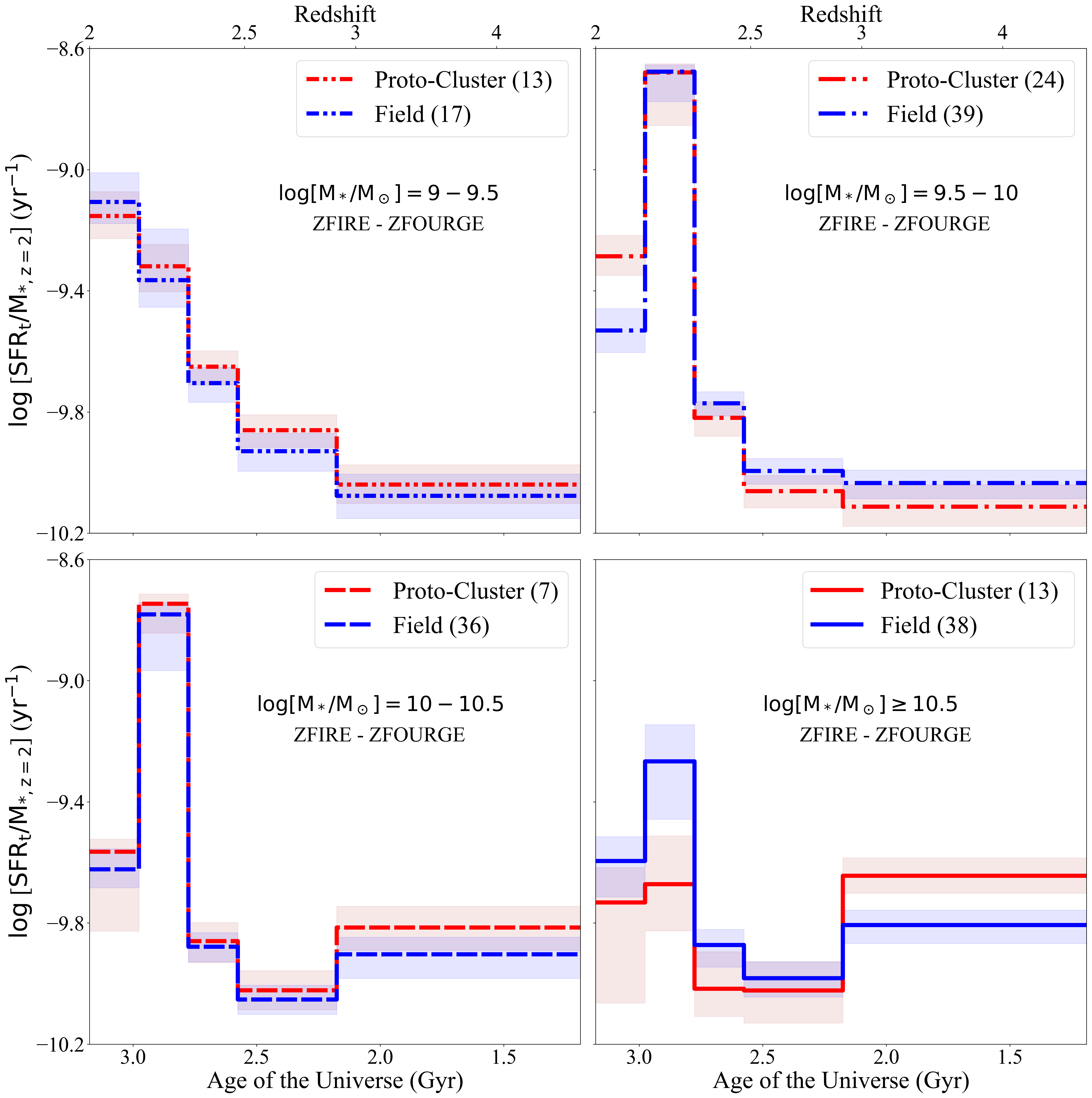}

    \caption{ Specific star formation rate ($\log(\rm{SFR}(t)/\rm{M}_*(t_0))$) histories from \zfire\ - \zfourge\ observations derived with \pros\ vs age of the Universe (bottom x-axis) and corresponding redshift (top x-axis) for field (blue) and proto-cluster (red) galaxies at $z\sim2$. The four panels correspond to 4 mass bins (from left to right): $9\leq$\logMstarMsun$<9.5$, $9.5\leq$\logMstarMsun$<10$, $10\leq$\logMstarMsun$<10.5$, \logMstarMsun$\geq 10.5$. The solid and dashed lines show the bootstrapped median sSFH and the shaded region shows the error in median. The number of galaxies in each bin is given in parenthesis in each label. The bottom right panel (\logMstarMsun$\geq 10.5$) shows evidence for the  suppressed star formation activity in massive proto-cluster galaxies in the recent time bins.   }
    \label{fig:sfh_pros}
\end{figure*}

\begin{figure*}
	\noindent
	\includegraphics[scale = 0.32]{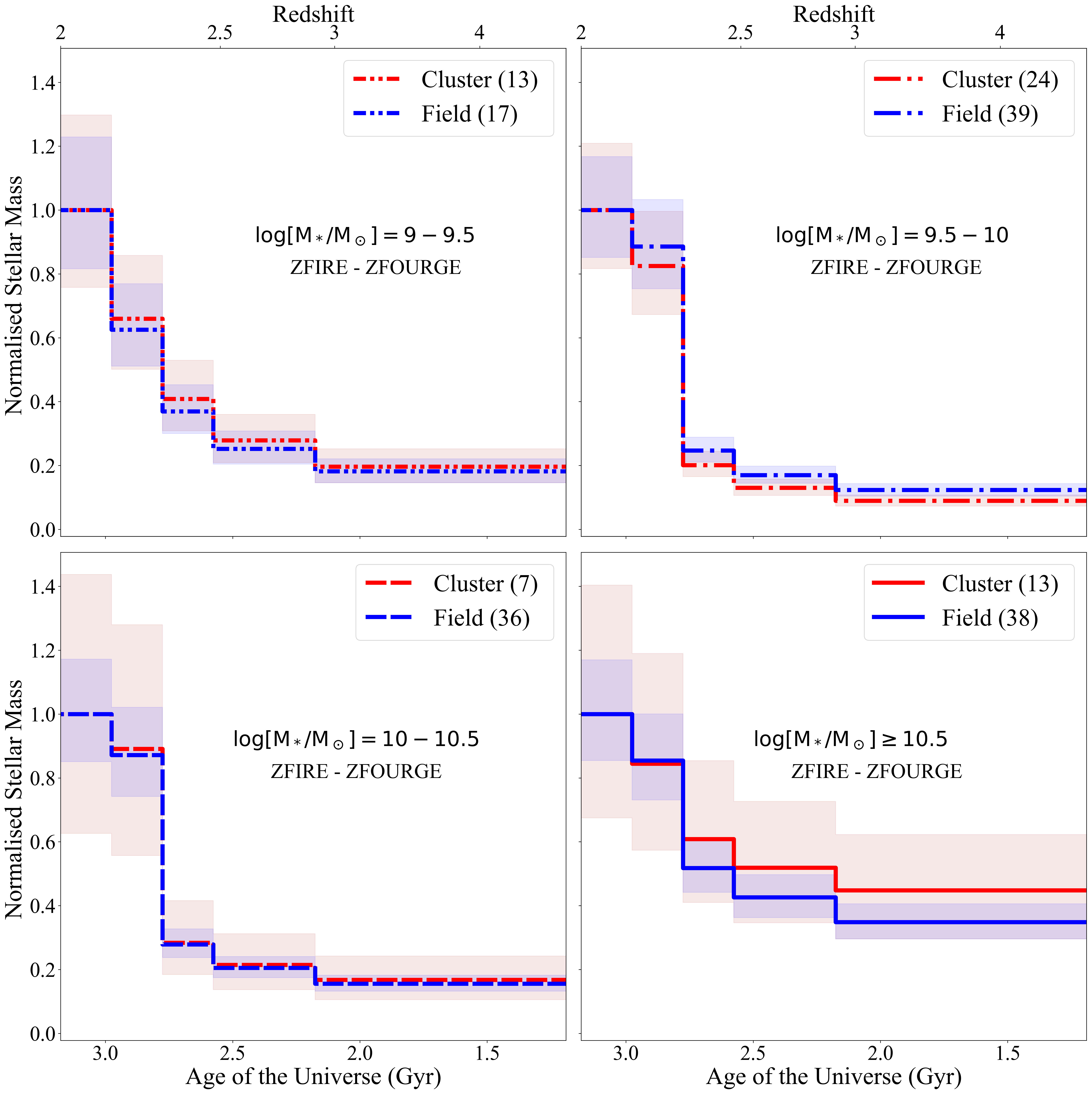}

	\caption{ Normalised cumulative stellar mass in galaxies from  \zfire\ - \zfourge\ observations derived with \pros\ vs age of the Universe (bottom x-axis) and corresponding redshift (top x-axis) for field (blue) and proto-cluster (red) galaxies at $z\sim2$. The four panels correspond to 4 mass bins (from left to right): $9\leq$\logMstarMsun$<9.5$, $9.5\leq$\logMstarMsun$<10$, $10\leq$\logMstarMsun$<10.5$, \logMstarMsun$\geq 10.5$. The solid and dashed lines show the bootstrapped median growth of stellar mass and the shaded region shows the $1 \sigma$ error in median. The number of galaxies in each bin is given in parenthesis in each label. The bottom right panel (\logMstarMsun$\geq 10.5$) shows evidence of early stellar mass buildup in massive proto-cluster galaxies compared to field galaxies. }
	\label{fig:stellarmass_pros}
\end{figure*}

\begin{figure*}
   \noindent
  
    \includegraphics[scale = 0.32]{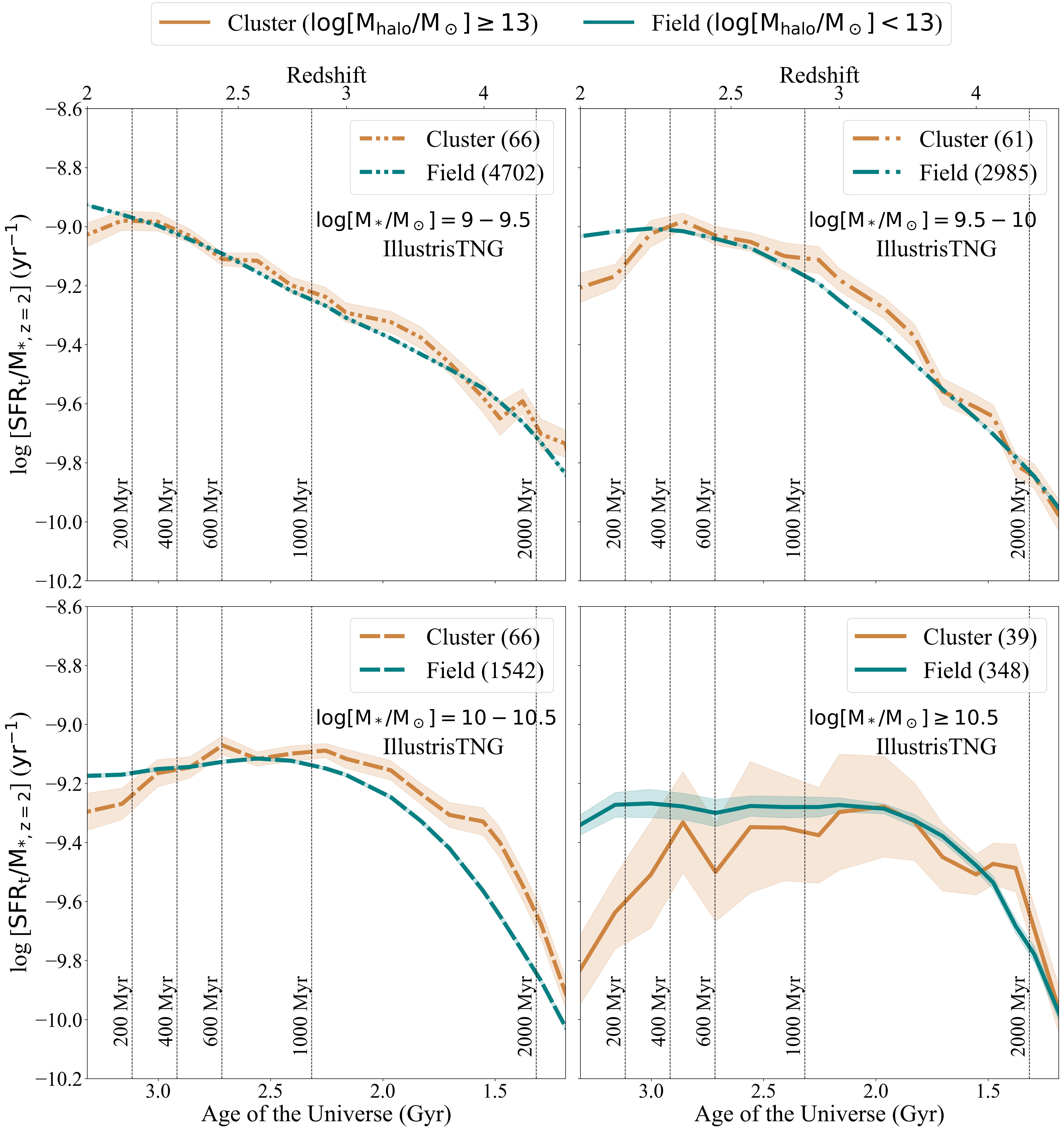}
     \caption{ Instantaneous specific star formation rate ($\log[\rm{SFR}(t)/\rm{M}_{*,z=2}]$) histories from \illustris\ vs age of the Universe (bottom x-axis) and corresponding redshift (top x-axis) for field (teal) and cluster (orange) galaxies at $z\sim2$. The four panels are the 4 mass bins (from left to right): $9\leq $\logMstarMsun$<9.5$, $9.5 \leq $\logMstarMsun$<10$, $10\leq $\logMstarMsun$<10.5$, \logMstarMsun$\geq10.5$. The solid and dashed lines show the bootstrapped median sSFH and the shaded region shows error in the median. The black dashed lines correspond to the age bins used in SFHs from observations (Figure \ref{fig:sfh_pros}). The number of galaxies in each bin is given in parenthesis in each label. Similar to observations, galaxies in the stellar mass bin $10.5\leq $\logMstarMsun$\leq11$ (bottom right panel) show effect of environment on the sSFH in contrast to lower mass bins where field and cluster galaxies have comparable sSFH. }
    \label{fig:sfh_illustris}
\end{figure*}

To study the effect of environment on the star formation of galaxies we compare the SFH of proto-cluster and field galaxies at $z\sim2$.  
Figure \ref{fig:sfh_pros} and \ref{fig:sfh_illustris} shows the star formation histories of galaxies in proto-cluster and field environment from \zfire\ -\zfourge\ observations and the \illustris\ simulations. 

\subsubsection{SFH Vs Environment: Observations}

We derive the star formation histories of galaxies at $z\sim2$ using the extensive \zfourge\ photometry and the SED fitting tool  \pros. Figure \ref{fig:sfh_pros} shows the median star formation rates in each age bin normalised to the total stellar mass of the galaxy at the $z\sim2$ of proto-cluster (red) and field (blue) galaxies. The effect of environment is most evident on the highest mass galaxies (\logMstarMsun $ > 10.5$). The same trend is not observed in the other mass bins although cluster galaxies in the \logMstarMsun $=10-10.5$ bin also show slightly ($\sim1\sigma$) higher star formation in the most recent age bin compared to the field sample (log sSFR = $-0.81\pm0.09\ \rm{Gyr}^{-1}$ in proto-cluster and $-0.92\pm0.06\  \rm{Gyr}^{-1}$ in field).

Massive proto-cluster galaxies show higher star formation in the earliest age bin ($>1$ Gyr in look-back time; until $\sim 2.1$ Gyr of age of the Universe) compared to field sample (log sSFR = $-0.65\pm0.06\,\rm{Gyr}^{-1}$ in proto-cluster and $-0.8\pm0.04\,\rm{Gyr}^{-1}$ in field) and a lower star formation (log sSFR = $-0.71\pm0.03\,\rm{Gyr}^{-1}$ in proto-cluster and $-0.61\pm0.02\,\rm{Gyr}^{-1}$ in field) in the more recent age bins ($0-200$ Myr in look-back time; $\sim 3.1 - 2.9 $ Gyr of age of the Universe). This indicates an earlier formation and stellar mass build up of massive proto-cluster galaxies compared to field galaxies. 

Figure \ref{fig:stellarmass_pros} shows the normalised cumulative median stellar mass build-up in proto-cluster (red) and field (blue) galaxies in the four stellar mass bins. The most massive  (\logMstarMsun $ > 10.5$) proto-cluster galaxies build up their stellar mass faster than the field galaxies, whereas proto-cluster and field galaxies with \logMstarMsun $ < 10.5$ have consistent stellar mass build-up history.  

Environmental effects arising from the interactions of galaxies with the ICM through processes like ram pressure stripping, starvation, harassment etc. are shown to be more effective in quenching lower mass galaxies, which are rarely quenched in field environment \citep{Medling2018}. In our sample, we do not see any effect of environment on the low mass galaxies (\logMstarMsun $ =9-9.5$). In the second stellar mass bin (\logMstarMsun $ =9.5-10$), we find a higher SFR in the the most recent age bin (3.1- 3.3 Gyrs in the  age of the universe) for proto-cluster galaxies than the field sample. Figure \ref{fig:stellarmass_pros} shows the difference in the stellar mass formed in the proto-cluster galaxies in the small time scale (200 Myrs) of the most recent age bin is not significant to be seen as a deviation from the otherwise closely following SFH of the  field galaxies. The star formation histories of star forming proto-cluster galaxies in the low mass bins closely follow that of the field galaxies in the same mass bin. Importantly, however, this could be due to the bias of \zfire\ sample towards star forming galaxies of SFR $>0.8$ \sfr\ \citep{yuan2014}.

\subsubsection{SFH Vs Environment: Simulations}
The star formation histories from \illustris \ (Figure \ref{fig:sfh_illustris}) track the evolution of instantaneous star formation of galaxy in twice the stellar half mass radius of the galaxy in cluster (orange) and field (teal) environment at $z = 2$. The instantaneous SFRs are different from the SFHs based on stellar ages from \pros, but can still be compared qualitatively. As the observational limit on H$\alpha$ flux biases our observational sample towards galaxies with SFR $>0.8 $ \sfr, we have imposed an analog SFR threshold to the simulated galaxies to have a comparable sample. In practice, we have imposed a SFR cut of $0.4$ \sfr\ on the selection of galaxies from \illustris\ to match the observations (refer to Section \ref{sec:illustris}).  

Figure \ref{fig:sfh_illustris} shows the star formation rate normalised to the stellar mass of cluster and field galaxies at $z=2$ from \illustris. Similar to observations, star formation in cluster galaxies declines with time in the high mass (\logMstarMsun $> 10.5$) sample, whereas in the low mass sample, field and cluster galaxies have comparable SFHs. We also see a slight elevation ($<0.2$ dex) of star formation in cluster galaxies compared to field galaxies in mass bin \logMstarMsun $= 10 - 10.5$. The average SFH of low mass galaxies (\logMstarMsun $< 10.5$) is comparable in cluster and field environments.

Contrary to our result of rising of star formation in low mass cluster galaxies, \cite{Donnari2021} find a quenched fraction of $\sim 0.4$ in the low mass ($9>$\logMstarMsun$>9.5$) satellite galaxies in \illustris\ at $z=2$. By selecting galaxies above SFR $> 0.8$ \sfr\ in observations and $> 0.4$ \sfr\ in simulations, we are probably removing galaxies whose star formation activity has already been suppressed. This is certainly the case for the simulated sample. For galaxies in the real Universe, we can only say that either the star formation in low mass galaxies is not affected by the environment, or that the environmental suppression of star formation happens rapidly in low mass cluster galaxies compared to high mass galaxies.

\subsection{Star Formation Histories and Stellar Mass}
\label{sec:SFHvssm}
\begin{figure*}
\noindent

\includegraphics[scale = 0.3]{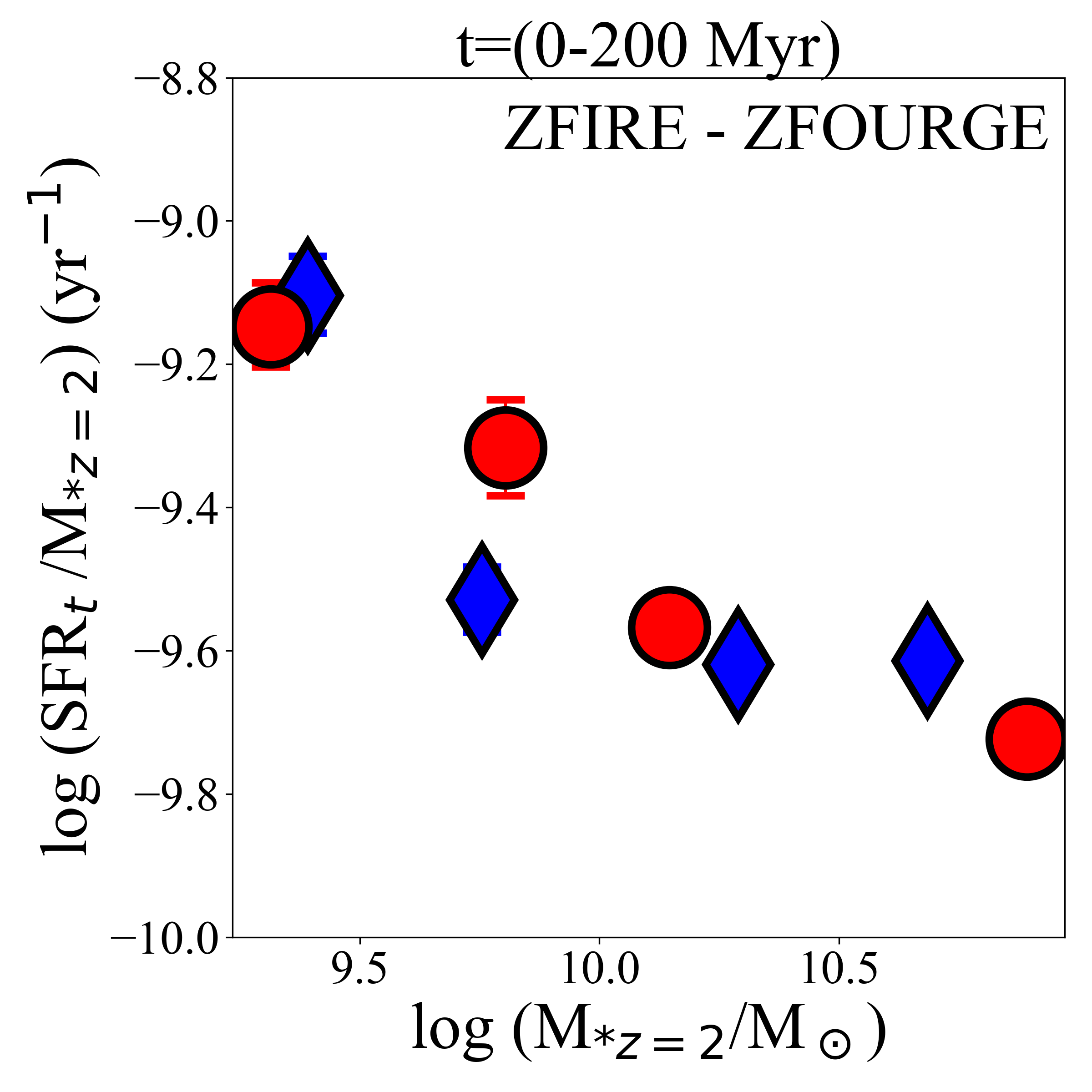}
\includegraphics[scale = 0.3]{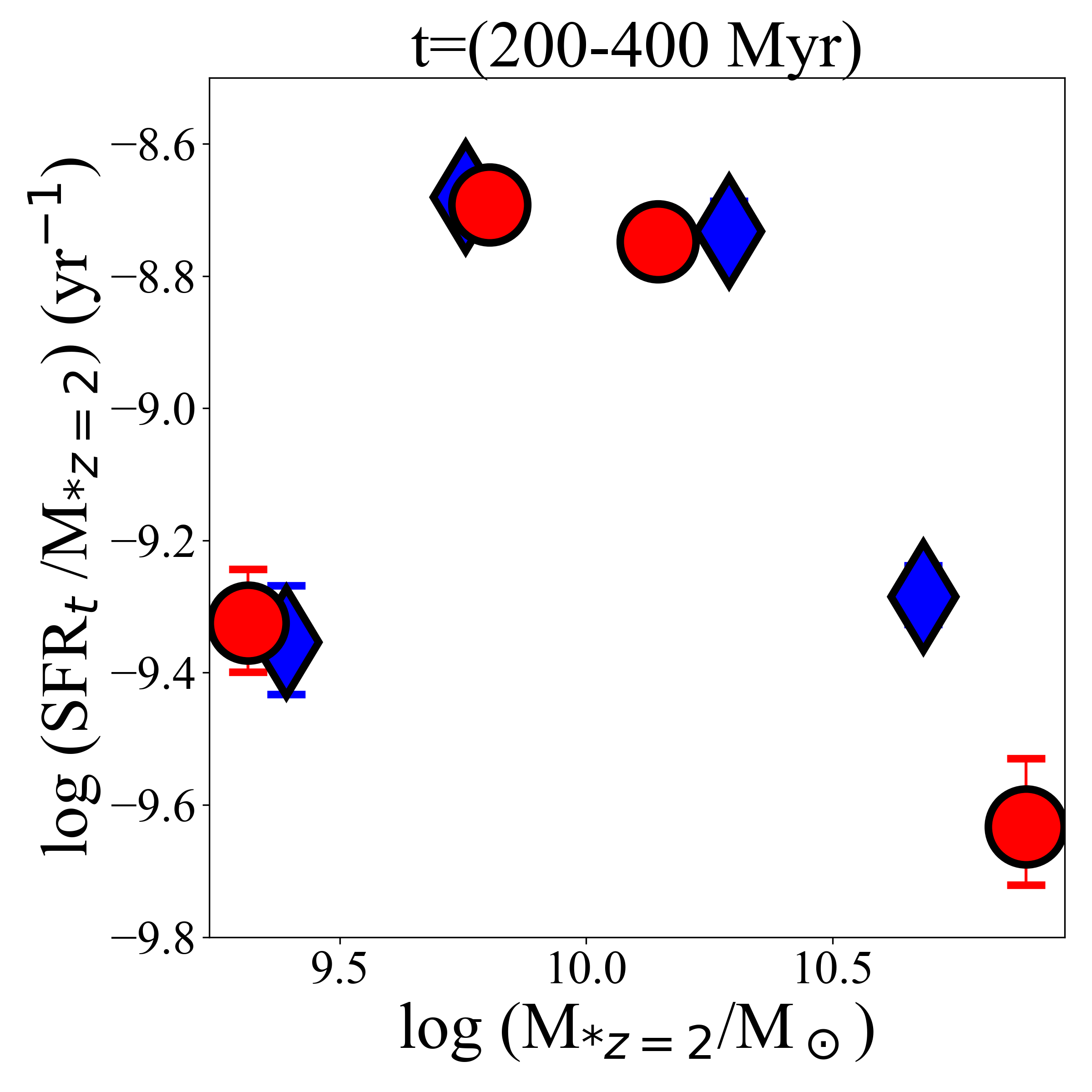}
\includegraphics[scale = 0.3]{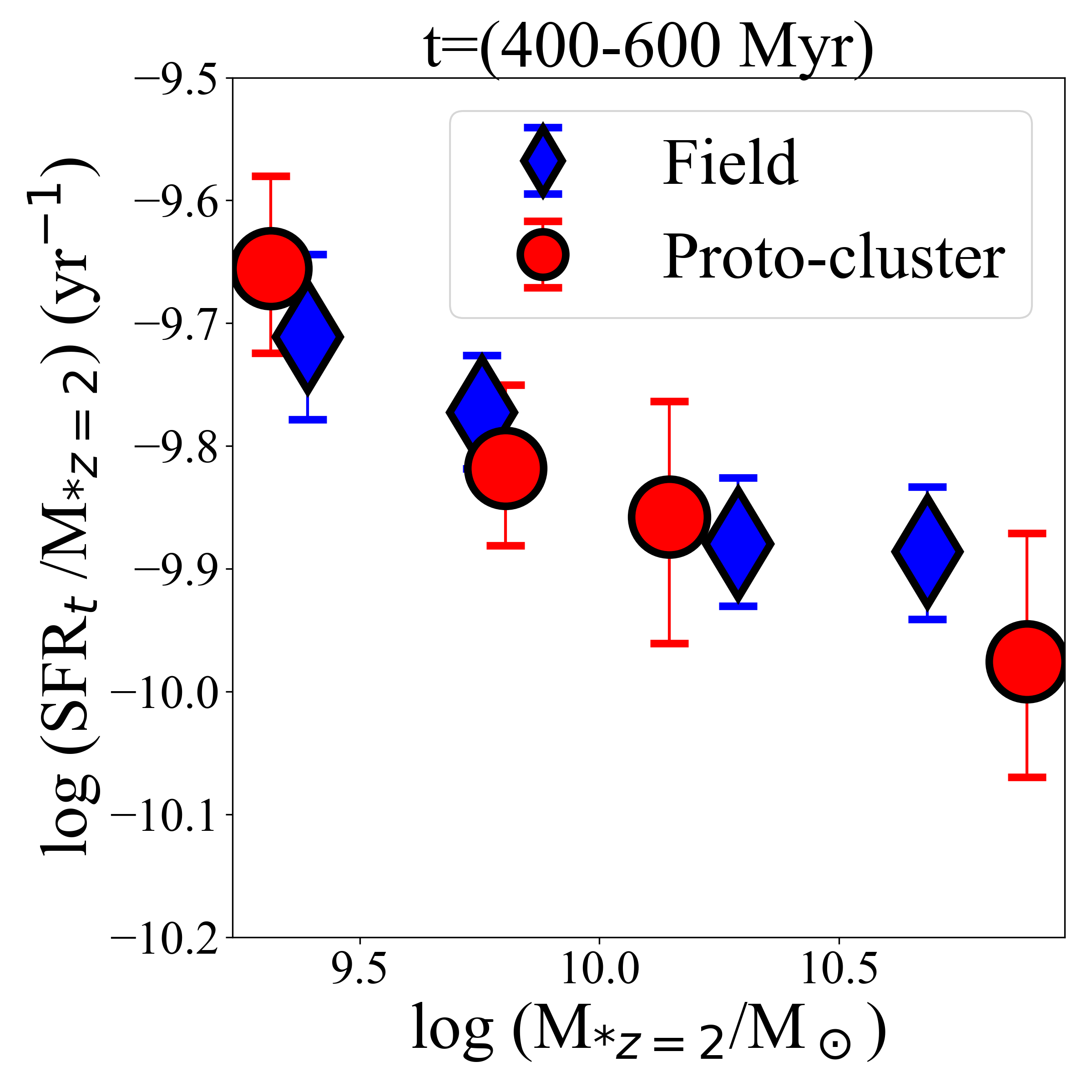}
\end{figure*}

\begin{figure*}
\noindent
\centering
\includegraphics[scale = 0.3]{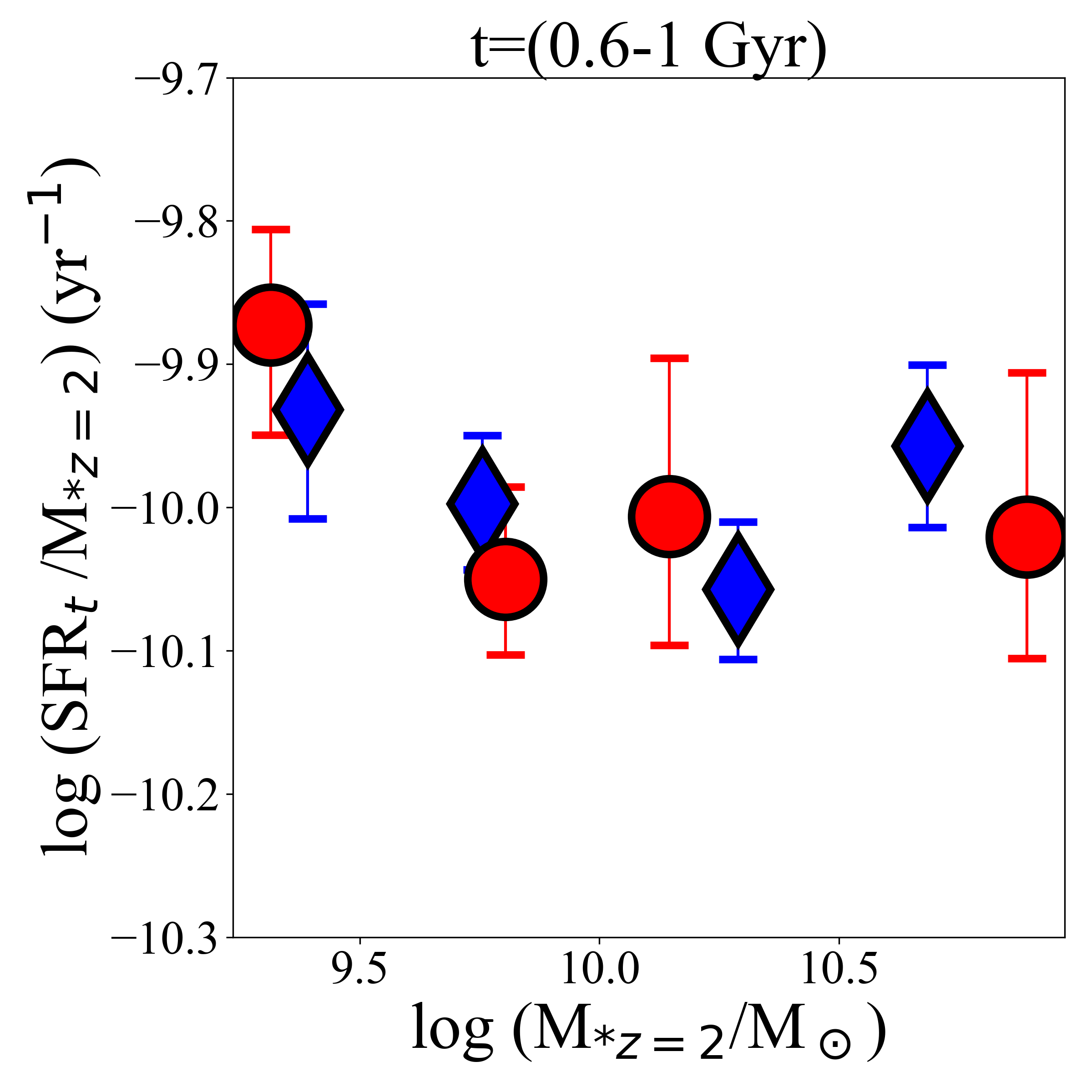}
\includegraphics[scale = 0.3]{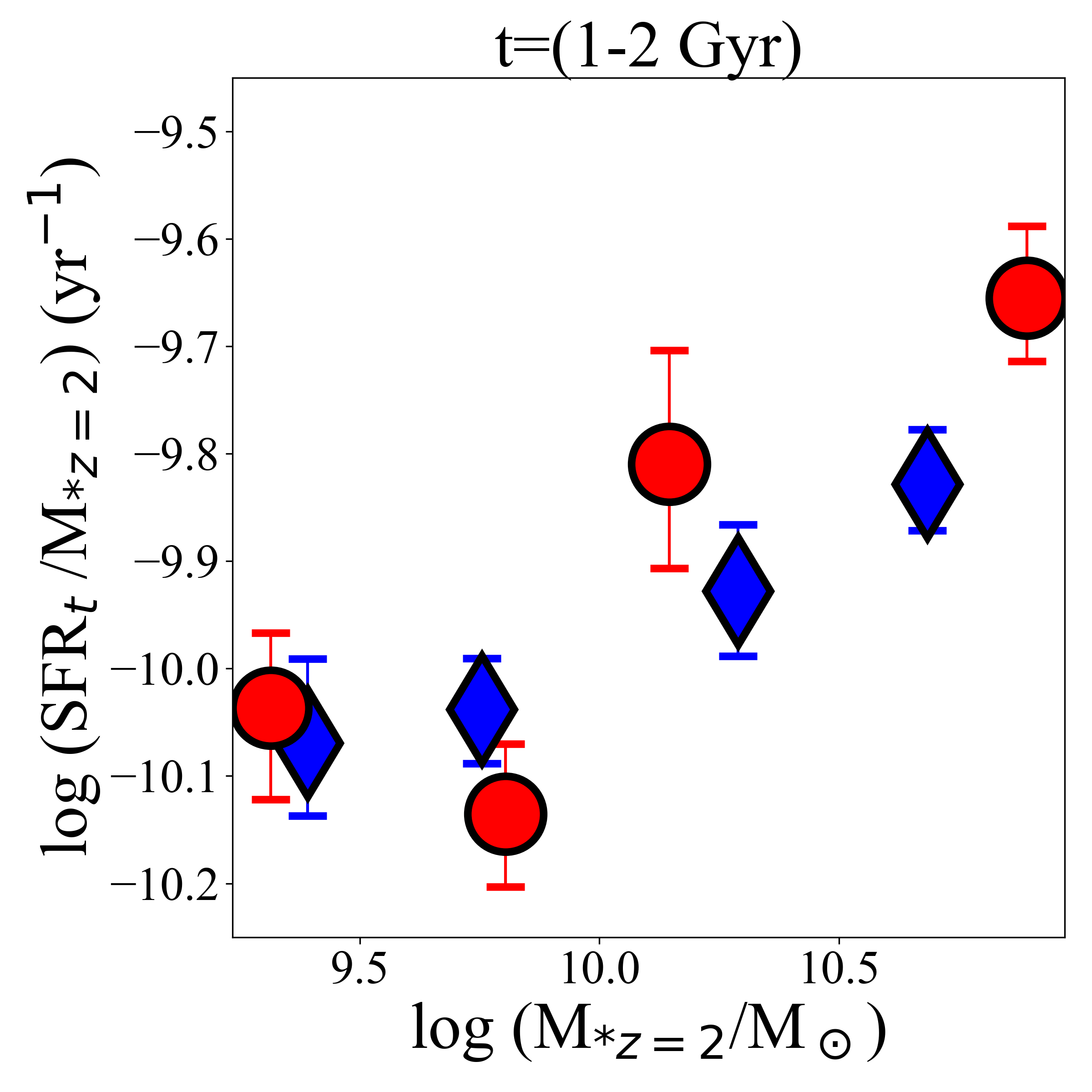}
 \caption{Specific Star formation rate ($\rm{SFR}(t)/\rm{M}_{*,z=2}$) from observations vs stellar mass in 5 look back time age bins: $0-200$ Myr, $200 - 400$ Myr, $400 - 600$ Myr, $600- 1000$ Myr and $>1$ Gyr (top-left to bottom right) for proto-cluster (red circles) and field (blue diamonds) galaxies. The sSFRs are plotted against the median stellar mass of each mas bin. The error bars represent the error in median. In the earliest age bin (t = 1-2 Gyr) massive galaxies (\logMstarMsun $= 10-11$) are forming more stars compared to lower stellar mass galaxies (\logMstarMsun $= 9-10$) in contrast to the most recent age bin (t = 0-200 Myr), where lower mass galaxies are forming more stars than the massive galaxies across environments.  }
    \label{fig:sfh_bins}
\end{figure*}

The star formation activity in a galaxy is closely related to its stellar mass at $z= 0-4$ \citep{Noeske2007,Tomczak2016,Koyama2018}. Since, SFH tracks how galaxies grow its stellar mass, the SFH of galaxies are also related to its stellar mass \citep{Thomas2005}. We find comparable results from the star formation histories derived from \zfourge\ photometry using \pros\ and \illustris.

\subsubsection{SFH Vs Stellar Mass: Observations}
Figure \ref{fig:sfh_bins} shows dependence of star formation rates in each age bin normalised to the stellar mass of the galaxy at the time of observation. From top left to bottom right, the age bins range from recent to earliest. The most massive galaxies (\logMstarMsun $> 10.5$) have higher sSFR ($\sim 0.6$ dex) compared to low mass galaxies in the earliest age bin, indicating that massive galaxies formed their stellar masses early on irrespective of the environment. Compared to that, towards the most recent age bin ($0-200$ Myr), we find this trend gradually reverses. In the most recent age bin ($0-200$ Myr, the most massive galaxies have  lower sSFR ($\sim 0.8$ dex) compared to the lowest mass galaxies.

From SED fitting of observations with \pros, figure  \ref{fig:stellarmass_pros} shows that the high mass cluster galaxies (\logMstarMsun $> 10.5$) form $\approx 45 \pm 8\%$ of stellar mass in the first $\sim2$ Gyr of the Universe compared to $9\pm 1\%$ to $19\pm2\%$ of stars formed for lower stellar mass galaxies in the same environment. In the field sample, the high mass galaxies (\logMstarMsun $ > 10.5$) form  $\approx 31 \pm 2\%$ of their stellar mass in the first $\sim 2$ Gyr of the Universe compared to $12\pm 1\%$ to $17\pm 1\%$ in the lower stellar mass galaxies. Figure \ref{fig:sfh_pros} also shows that the shape of the SFH for galaxies is dependent on the stellar mass of the galaxies. The highest mass bin have a constant SFH compared to galaxies in the lowest mass bin, for which the median SFH is rising. Whereas galaxies in stellar mass range \logMstarMsun $= 9.5-10.5$ have a bursty star formation history. 

The rising SFH is measured only in the lowest stellar mass galaxies, where the observational limitation of our sample excludes low stellar mass galaxies with low star formation rates. The low stellar mass galaxies that have suppressed star formation either due to environmental or secular processes are thus not included in our sample and this would explain the rising SFH measured in the lowest stellar mass bin for our sample. 

\subsubsection{SFH Vs Stellar Mass: Simulations}
Figure \ref{fig:sfh_illustris} shows the star formation rates of the cluster and field galaxies from \illustris. The SFHs of TNG100 galaxies show similar trends to the observations. The SFH of high stellar mass  galaxies (\logMstarMsun $> 10$) either plateaus with time in the field environment or drops in clusters. On the other hand, low stellar mass galaxies (\logMstarMsun $ < 10$) have rising SFHs irrespective of the environment. This result is comparable to our observations (Figure \ref{fig:sfh_pros}). The SFH of galaxies from \illustris\ in the highest stellar mass bin (\logMstarMsun $ > 10.5$) plateaus $\approx 500$ Myr  earlier than the galaxies in mass bin  ($10> $ \logMstarMsun $> 10.5$) and $\approx 1$ Gyr  earlier than the galaxies in mass bin  ($9.5>$ \logMstarMsun $> 10$). Massive galaxies (\logMstarMsun $> 10.5$) form most stars ($38 \pm 3 \%$ and $32 \pm 0.4 \%$ of their total stellar mass in proto-cluster and field environments, respectively) in the first $2$ Gyr.\\

Our analysis indicates an earlier formation and evolution of massive galaxies irrespective of environment compared to lower mass galaxies. This result is in agreement to many theoretical and observational studies \citep{Thomas2005,SanchezBlazquez2006, Renzini2016}. Recent observational study by \cite{Webb2020} have comparable results for quenched galaxies at $z<1.5$. \cite{Thomas2005} show that most massive early type galaxies at $z=0$ have peak star formation activity $1-2$ Gyr before low stellar mass galaxies, comparable to our results.

\subsection{Stellar Age: Mass and Environment}
Our results from \zfourge\  and \illustris\ show significant effect of environment on the SFH of high mass galaxies at $z=2$. In this Section, we explore in \illustris\ if earlier formation or merger histories could explain the measured decline in star formation activity of massive cluster galaxies. 

\begin{figure}
\noindent
\centering
\includegraphics[scale = 0.28]{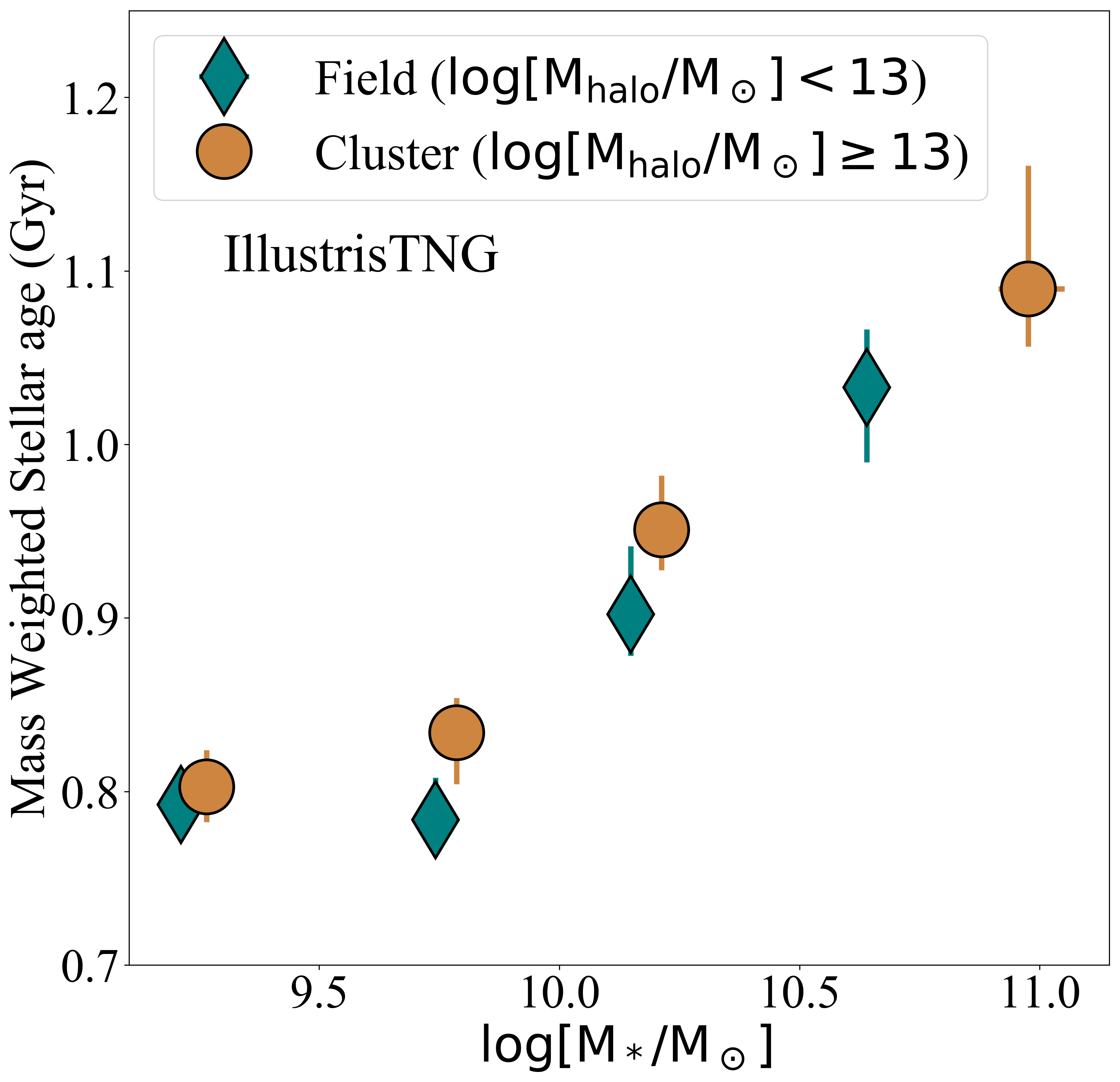}

 \caption{Mass weighted stellar age (Gyr) vs log stellar mass ($\rm{M}_\odot$) of the galaxy in cluster(orange) and field(teal) environments at $z=2$ in \illustris. The massive galaxies (\logMstarMsun $>10.5$) are $\approx$190 Myr older than the low mass galaxies (\logMstarMsun $=9-9.5$). Stellar population in cluster galaxies have comparable ages to the field galaxies across the stellar mass range.}
    \label{fig:age}
\end{figure}

To estimate if massive galaxies in the proto-cluster environment formed earlier than the field galaxies, we analyse the average age of their stellar populations in \illustris. We extract the mass weighted stellar age for each galaxy in our field and cluster environment as follows:
\begin{equation}
    t_g = \frac{\sum_{i=1}^{N} t_iM_i}{\sum_{i=1}^{N}M_i}
\end{equation}
where $t_g$ is the mass weighted stellar age of the galaxy, $t_i$ and $\rm{M}_i$ are the age and mass of each stellar particle in the galaxy and N is the total number of stellar particles in the galaxy. We use the snapshot particle data of \illustris\  \citep{Nelson2019a} to get the stellar ages of galaxies at $z=2$. Figure \ref{fig:age} shows the mass weighted stellar ages of the cluster (orange) and field (teal) galaxies in the respective stellar mass bins. The error bars show the error in the median of the distribution.

The median mass weighted stellar ages increase with the stellar mass of the galaxy in the cluster. The mass-weighted stellar ages of cluster galaxies is comparable to those of the field galaxies across the stellar mass ages. Our results do not change significantly when comparing the ages of the oldest stellar particle in the galaxy.



The median stellar ages of the lowest mass galaxies (\logMstarMsun $ =9-9.5$) is $\approx 800\pm 20$ Myr and high mass galaxies (\logMstarMsun $>10.5$) is $\approx 980 \pm 10$ Gyr. The highest mass galaxies are $\approx 190 \pm 30$ Myr older than the low stellar mass galaxies. The cluster galaxies are the same age as the field galaxies across the stellar mass bin. This result is different to the observational study by \cite{Webb2020}, who find that at $z=1$, cluster galaxies are $310$ Myr older than the field galaxies in the mass bin  \logMstarMsun $=10-11.8 $.

\subsection{Merger events}
Many theoretical and observational studies correlate galaxy mergers with stellar mass growth, increased gas fractions, enhanced AGN activity and enhanced SFR \citep{Kewley2006,Ellison2015,Dutta2019,Moreno2019, Hani2020}. \cite{Watson2019} show that galaxy mergers are twice as frequent in the proto-cluster environment compared to the field at $z\sim2$. Using \illustris\, \cite{Hani2020} show that the enhancement of the SFR activity due to mergers correlates with the stellar mass, mass ratio and the gas fraction of the merging pair. The decay of SFR enhancement in post merger phase happens over $500$ Myr and the galaxies that underwent strongest merger driven starburst events quench on a faster timescale. %

\begin{figure*}
\noindent
\centering
    \includegraphics[scale = 0.31]{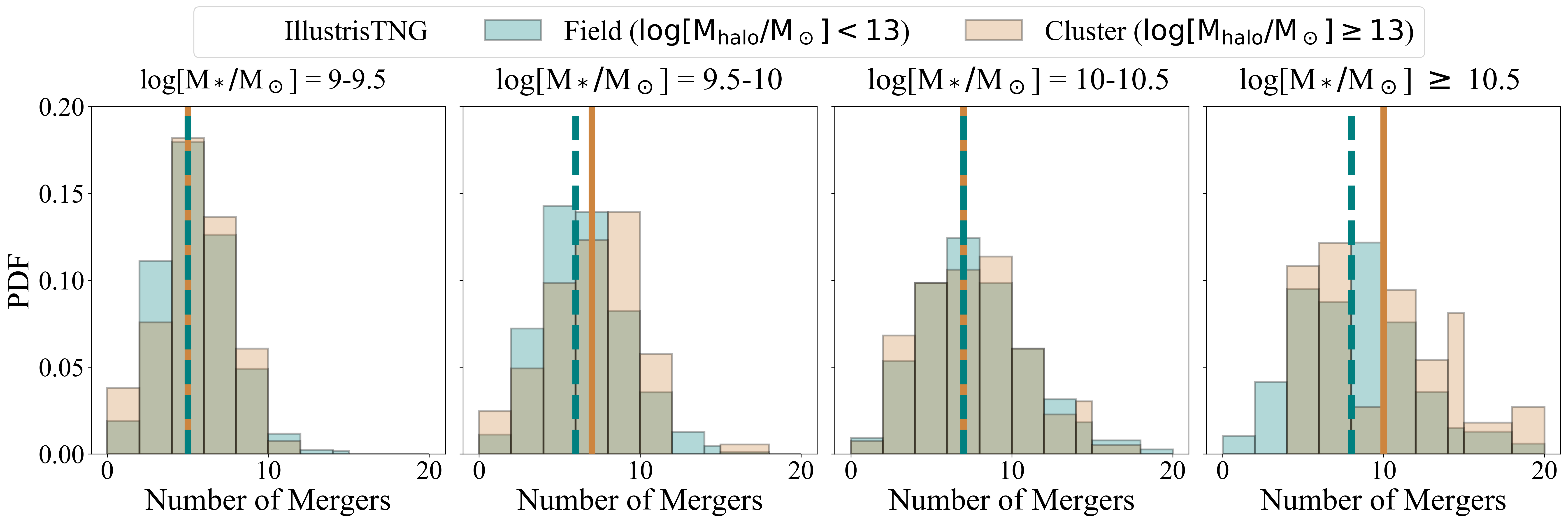}
 \caption{Distribution of total number of mergers (mass ratio $> 0.1$) in field (teal) and cluster (orange) galaxies in four stellar mass bins in \illustris\ at $z=2$. (left to right: \logMstarMsun $ = 9-9.5,\ 9.5-10,\ 10-10.5,\ \geq 10.5$ ). Massive galaxies (\logMstarMsun $>10.5$) on average have experienced 3-5 more mergers (mass ratio $> 0.1$) compared to the low mass galaxies (\logMstarMsun $= 9-9.5$) in both cluster and field environment.  }
    \label{fig:mergerdist}
\end{figure*}

We explore if the differences in the SFH of massive galaxies between environments is driven by mergers. We use the merger catalogs by \cite{Rodriguez-Gomez2015} from \illustris\ to explore the effect of merger on the star formation histories. Figure \ref{fig:mergerdist} shows that the distribution of total mergers (mass ratio $> 0.1$) encountered in cluster (orange) and field (teal) environment in different stellar mass bins is comparable. However, massive galaxies (\logMstarMsun $> 10.5$) on average have experienced $8\pm0.3$ and $10\pm1$ mergers (mass ratio $> 0.1$) in field and cluster sample respectively, compared to $5\pm0.3$ mergers experienced by low mass cluster and field galaxies (\logMstarMsun$ = 9-9.5$) in their lifetimes. 

\begin{figure}
\noindent
\centering

\includegraphics[scale = 0.27]{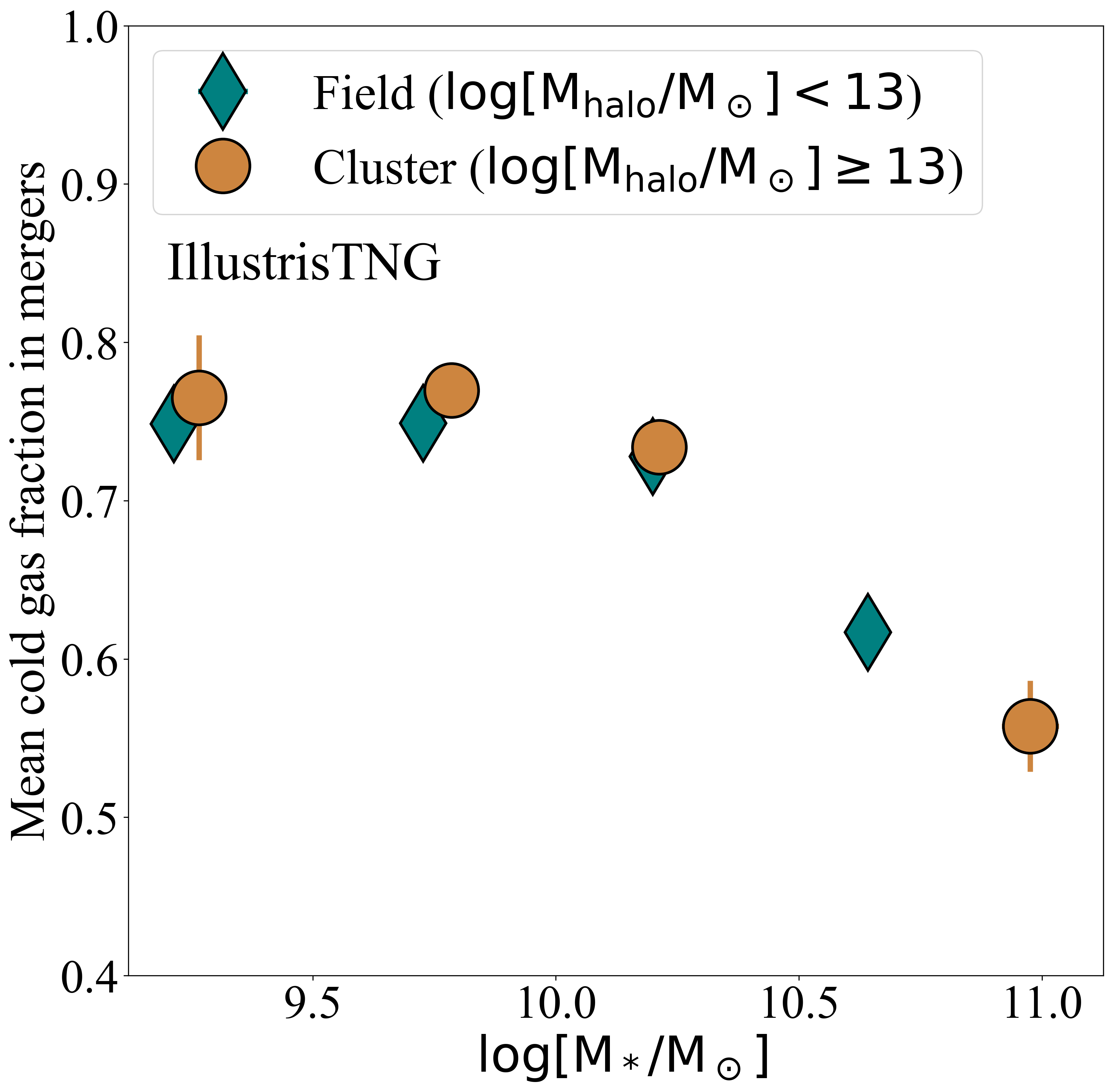}

 \caption{Mean cold gas fraction in mergers Vs stellar mass of galaxies in field (teal diamonds) and cluster (orange circles) environment. The markers show the median and error bars are error in median for the sample. The mean cold gas fraction of mergers decreases with increasing stellar mass of the galaxies and are comparable across environments. }
    \label{fig:coldgasfrac}
\end{figure}

While merger events often lead to further star formation, gas poor mergers do not and in turn will not affect the average stellar age of the galaxy. To estimate if massive galaxies have experienced gas poor mergers we analyse the cold gas fraction and total gas fraction of the mergers. We use the merger history catalogs by \cite{Rodriguez-Gomez2017} to get the mean cold gas fraction (weighted by the stellar mass) of all the progenitors until $z=2$ (Figure \ref{fig:coldgasfrac}). The mean cold gas fraction of cluster and field galaxies are comparable across stellar mass bins.

To calculate if the difference in SFHs could be driven by the total supply of gas, we calculate the total gas fraction of the progenitors for galaxies at each redshift snapshot as: 

\begin{equation}
    f_{gas,z} = \frac{\sum_{p=1}^{N} \rm{M_{gas}}_{p, z}}{\sum_{p=1}^{N} \rm{M_{gas}}_{p, z}+\sum_{p=1}^{N} \rm{M_{*}}_{p, z}}
\end{equation}

where $f_{gas,z}$ is the total gas fraction of all progenitors (p) at redshift $z$, $\rm{M_{gas}}_{p, z}$ and $\rm{M_{*}}_{p, z}$ are the total gas mass and stellar mass of each progenitor $p$ at the same redshift snapshot $z$.

\begin{figure}
\noindent
\centering

\includegraphics[scale = 0.27]{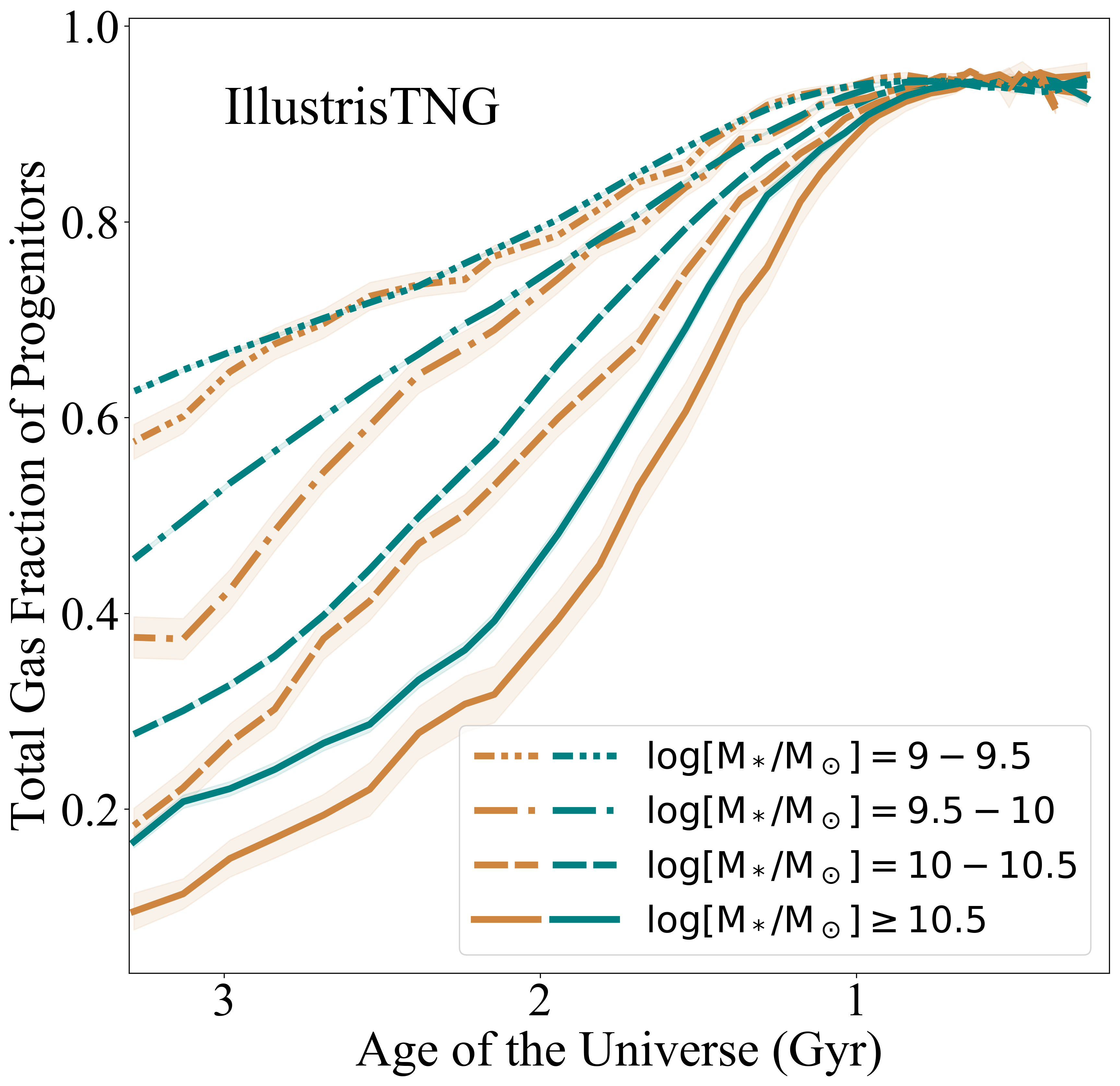}

 \caption{Total gas fraction of progenitors Vs age of the Universe for four stellar mass bins in field and cluster environment. Progenitors of the most massive cluster galaxies (\logMstarMsun $\geq 10.5$, solid orange line) are consistently gas poorer in comparison to the progenitors of field galaxies in the same stellar mass bin (teal solid line) since $\sim1$ Gyr of the age of the Universe.}
    \label{fig:gasfrac}
\end{figure}

Figure \ref{fig:gasfrac} shows the total gas fraction of the progenitors for the cluster (orange) and field (teal) galaxies in the 4 stellar mass bins. The cluster galaxies in the highest stellar mass bin (\logMstarMsun$ >10.5 $) encounter lower gas fractions mergers consistently since 1 Gyr after the Big Bang compared to field galaxies in the same stellar mass bin. On the other hand, the total gas fraction of progenitors in the lower stellar mass bin (\logMstarMsun$ = 9-9.5 $) are comparable across environments throughout their merger histories.



\section{Discussion}
\label{sec:discussion}
The star formation rate of a galaxy depends strongly on its stellar mass and local environment. This dependence has been demonstrated to evolve since $z\sim2$ in both observations and cosmological simulations \citep{ Peng2010, Muzzin2011, Tran2015, Tomczak2016, Kawinwanichakij2017, Darvish2017, Donnari2019, Donnari2020a, Donnari2021, Webb2020}. In this work, we have compared the star formation histories of star-forming galaxies in the cluster and field environments at $z\sim2$ using observations and simulations. 


\subsection{Star Formation History: Environment and Stellar Mass}
In Section \ref{sec:results} we have shown that the star formation histories of galaxies from \zfourge\ at $z\sim2$ strongly depends on the stellar mass of the galaxy in both proto-cluster and field environments. Figure  \ref{fig:stellarmass_pros} shows that galaxies in the highest mass bin have formed $45 \pm 8 \%$ and $31 \pm 2 \%$ (in proto-cluster and field environments) of their present stellar mass in the first $\sim 2$ Gyr. Compared to the high mass galaxies in the same time epoch (first $\sim 2$ Gyr), the lowest mass galaxies form $<20\%$ of their present stellar mass in both proto-cluster and field environments.  The star formation histories from \pros\ also show that the most massive galaxies have a constant or declining star formation history as opposed to the rising star formation history of the lowest mass galaxies in both high and low density environments. 
This points to a faster and early stellar mass build up in the most massive galaxies and a delayed evolution of the lowest mass galaxies.

This result is consistent with the observational and theoretical studies that show a mass dependence of star formation histories \citep{ Thomas2005,Poggianti2006, Sanchez-Blazquez2009,Thomas2010, Webb2020}. These studies show that the evolution of more massive galaxies happen over shorter time scales compared to their lower mass counterparts. Stellar population studies predict a difference of $\sim 2 $ Gyr between the evolution of high mass cluster early type galaxies and high mass field early type galaxies at $z\sim0$ \citep{Thomas2005, Renzini2006}, which is over estimated compared to our results. However, the $2 $ Gyr difference could be a result of redshift evolution. 

In our sample, the effect of environment on SFH of star forming galaxies in $z\sim2$ proto-cluster is present only in the highest mass galaxies (\logMstarMsun $ > 10.5$). he massive proto-cluster galaxies, which form $\approx 45 \pm 8\%$ of their stellar mass in the first 2 Gyr, have a declining star formation history compared to the massive field galaxies which forms $\approx 31 \pm 2 \%$ of their stellar mass in the first $\approx 2$ Gyr (Figure  \ref{fig:stellarmass_pros}). The field galaxies takes $\approx 2.8 $ Gyr to form $46\%$ of its total stellar mass. This shows a slower and delayed (by $0.8$ Gyr) evolution of high mass field galaxies compared to the proto-cluster galaxies. The lack of environmental effect on the SFH of low mass galaxies at $z\sim2 $ is comparable to the result in \cite{Papovich2018}, who find that the environmental quenching efficiency of galaxies decreases with stellar mass until $z\sim 0.5 $. Our conclusion for the observed low mass galaxies remains to be confirmed with an observational sample with SFR completeness below SFR threshold of $0.8$ \sfr.

We find similar trends of star formation histories in consistently-selected galaxies from \illustris\ (Figure \ref{fig:sfh_illustris}). As we go from the highest mass galaxies to the lowest mass galaxies we see that the plateauing of star formation occurs earlier for high mass galaxies compared to the low mass sample. For the highest mass galaxies (\logMstarMsun $= 10.5 - 11 $) the star formation plateaus at $\sim 1.5$ Gyr compared to  $\sim 2$ Gyr, and $2.7$ Gyr for galaxies in \logMstarMsun$ = 10 - 10.5 $ and \logMstarMsun$ = 9.5 - 10 $ mass bins respectively. Galaxies in the lowest mass bin (\logMstarMsun$ = 9 - 9.5 $) have a rising star formation until $z=2$. This also indicates early evolution of massive galaxies comparable to our results from \pros. We also find that the environment significantly affects the star formation histories of galaxies in the highest mass bin, with cluster galaxies forming more stars early on compared to the field galaxies similar to our results from observations.\\

In \illustris, the stellar mass formed by the most massive galaxies in the first 2 Gyr of the universe is $38\pm3\%$ in the cluster and $32\pm0.4\%$ in the field.  The difference in fraction of stellar mass formed in the first 2 Gyr of the universe is within $1\sigma$ error ($6\pm3.4\%$) compared to the observations ($14\pm10\%$). Cluster and field galaxies in the lowest mass bin form $24 \pm 0.7\%$ and $23 \pm 0.5 \%$ of their total stellar mass at $z=2$ which is lower than the fraction of total stellar mass formed in high mass galaxies, similar to our results from observations.\\

In \illustris, \cite{Donnari2020a} show that the fraction of quenched satellite galaxies at $z=2$ is $\sim 0.2-0.4$ in the stellar mass range \logMstarMsun $= 9 - 9.5 $ and \logMstarMsun $ = 10.5 - 11$. Due to the SFR cut imposed on the selection of galaxies from \illustris\ to match the observations, we do not see the suppression of SFH in low mass cluster galaxies. The low mass cluster galaxies experiencing the effect of environment and undergoing suppression of star formation would thus be removed due to the SFR threshold. We need spectroscopic redshift confirmation of faint ($K_{AB}>24$) galaxies at $z\sim2$ to be able to measure the effect of environment on SFH of the low mass quenched galaxies.

\subsection{Early Formation?}
Our results from the observations (\zfire\ - \zfourge) and \illustris\ simulations show signs of the onset of the star formation quenching in the most massive galaxies in the cluster environment. Using the \illustris\ simulations, we investigate a possible earlier formation of cluster galaxies driving the difference in SFHs (\ref{sec:SFHvsenv} and \ref{sec:SFHvssm}). Galaxies in the most massive bin are $\approx 190 \pm 30$ Myr older than lowest mass galaxies. Our result is unaffected if we consider the age of oldest stellar particle in the galaxy age as a proxy for formation time instead of the mass weighted stellar ages. 

The age difference of  $\approx 190 \pm 30$ Myr at $z=2$ between high and low stellar mass galaxies is consistent with the observational results at redshift $z\sim1$ where cluster galaxies are $\sim 300 - 400$ Myr older than the field galaxies \citep{VanDokkum2007,Webb2020} and $\sim 1$ Gyr at $z\sim0.1$ \citep{Thomas2005}. The increasing difference in stellar ages could be driven by the redshift evolution of ages of cluster and field galaxies. Nevertheless, the mass weighted stellar ages are unable to explain the measured difference in SFH of massive galaxies in cluster and field in our sample. This indicates that the suppression of star formation in high stellar mass cluster galaxies at $z=2$ in \illustris\ is not a result of earlier formation and evolution.

\subsection{Role of Mergers}

Studies have shown that massive galaxies grow their stellar mass through mergers \citep{Rodriguez-Gomez2015,Pillepich2018b,Gupta2020}. Recent work by \cite{Hani2020} uses \illustris\ to find that mergers enhance the SFR of post merger galaxies, but the relative increase in SFR depends on the stellar mass, mass ratio of the progenitor pair and the gas fraction of the progenitors. We investigate the possible role of mergers in shaping the SFHs. 

We track the merger histories of our sample from \illustris\ and find that in the combined cluster+field sample, on average massive galaxies (\logMstarMsun $\geq 10.5$) experience $8\pm0.26$ mergers compared to  $5\pm0.03$ mergers encountered by the low mass galaxies (\logMstarMsun = $9-9.5$) in $2\leq z < 20$ (Figure \ref{fig:mergerdist}). We speculate that the higher number of mergers experienced by massive galaxies early on leads to the build up of stellar mass and  higher SFR in the earlier time bins. The higher SFRs of massive galaxies in the early time bins (Figures \ref{fig:sfh_pros}, \ref{fig:sfh_illustris}) could lead to the depletion of gas reservoir faster compared to the lower mass galaxies (Figure \ref{fig:gasfrac}). The decreasing mean cold gas fraction of mergers with stellar mass of the galaxy also indicates the depletion of star formation fuel in the massive galaxies compared to the low stellar mass galaxies, explaining the relatively flat SFHs of massive galaxies (Figure \ref{fig:sfh_bins}).

The effect of environment on the SFHs of galaxies is only evident in the highest mass bin (\logMstarMsun $\geq 10.5$) in both observations (Figure \ref{fig:sfh_pros}) and simulations (Figure \ref{fig:sfh_illustris}). The high mass cluster galaxies experience more mergers ($10\pm1$) compared to the high mass field galaxies ($8\pm0.3$ mergers). Moreover, the progenitors of the massive cluster galaxies at $z=2$ have lower total gas fraction than the progenitors of high mass field galaxies even when the universe was 1 Gyr old (Figure \ref{fig:gasfrac}). Although, the mean cold gas fraction of the progenitors of two population is comparable at $z=2$ (Figure \ref{fig:coldgasfrac}). 

We hypothesize that the observed suppression of sSFR in massive cluster galaxies in the recent time bins is a delayed effect of the lower gas fraction in its progenitors. In other words, we find that massive galaxies in proto-clusters at $z\sim2$ show signatures of environmental effects not because of direct environmental processes due to their interaction with other cluster galaxies or the intra-halo medium, but rather because of the very nature of the environment they live in, which in turn affects their merger history and their opportunities to acquire gas.

Star formation rapidly progresses in the massive cluster galaxies with the available gas reservoirs. However, by $z=2$ massive cluster galaxies are starved because recent mergers have been systematically gas poorer in comparison to the massive field galaxies. The early onset of depletion of the gas reservoir in the massive cluster galaxies would cause the suppression of star formation by $z=2$ and could lead to an eventual quenching in the future via starvation. The environment-dependent depletion of gas fractions progresses from massive to low mass galaxies as we approach $z=2$ (Figure \ref{fig:gasfrac}). This suggests that the massive cluster galaxies would grow via dry mergers in the low redshift universe \citep{Tran2005, Webb2015}, and the environmental quenching would progress from massive galaxies to low mass galaxies in the cluster environment as is observed in the low redshift universe \citep{Donnari2021}.   

In a recent work \citep{Gupta2021} find that in \illustris\ star formation quenching in massive galaxies depends on their stellar size and is driven by the black hole feedback \citep{Davies2020,Zinger2020}. In future, we will test if the signs of early onset of star formation suppression in massive cluster galaxies is imprinted on the size of their stellar disks. We will further investigate if the growth and feedback of central super massive black hole is affected by the local environment of the galaxy.


\section{Summary}
In this paper, we have presented the first measurements of the SFHs of galaxies in the proto-cluster environment at $z=2$ using the \zfire\ - \zfourge\ surveys. We have compared our results with the SFHs of galaxies in different environments in the \illustris\ simulations and used the latter to provide a possible physical interpretation of the our findings. Our main results are summarised as:

1. \zfire\ - \zfourge: The SFHs of massive star forming galaxies ($10.5\leq $\logMstarMsun$ \leq 11$) in the proto-cluster are constant compared to the field galaxies in the same mass bin (Figure \ref{fig:sfh_pros}). In the first two Gyr of age of the Universe, massive proto-cluster galaxies form $45\pm 8 \%$ of total stellar mass compared to $31\pm 2 \%$ formed in massive field galaxies (Figure  \ref{fig:stellarmass_pros}). However, a similar dependence of SFHs on environment is not observed in galaxies in the lower mass bin (\logMstarMsun$ \leq 10.5$). 

2. \zfire\ - \zfourge: High mass galaxies form more  stars ($45\pm 8 $ to $31\pm 2\%$ in cluster and field environment respectively) in the earliest age bin ($> 2$ Gyr age of Universe), compared to low stellar mass galaxies ($17\pm 1\%$ to  $19\pm 2 \%$ in cluster and field environment respectively) in the same age bin (Figure  \ref{fig:stellarmass_pros}). This indicates a faster/earlier stellar mass build up of massive galaxies. 

3. \illustris: SFHs from simulations are comparable to our results from observations. The effect of environment is most prominent in the most massive galaxies (Figure \ref{fig:sfh_illustris}). Star formation in most massive cluster galaxies is suppressed compared to the field galaxies in the same mass bin. However, the SFHs of low mass galaxies do not show any dependence on environment. 

4. Stellar Ages: In \illustris, low mass galaxies are on average $190$ Myr younger than the high mass galaxies. However there is no difference in stellar ages in different environments across the studied stellar mass range (Figure \ref{fig:age}). Hence, the observed differences in the SFHs between cluster and field massive galaxies cannot be a result of early formation or earlier evolution of massive cluster galaxies.

5. Mergers and Gas Fractions: Based on the outcome of \illustris, we find that massive galaxies on average have experienced more mergers than low mass galaxies ($5$ mergers), irrespective of their environment (Figure \ref{fig:mergerdist}) by $z=2$. The mean cold fractions in mergers decrease with increasing stellar mass but are comparable across environments (Figure \ref{fig:coldgasfrac}). On the other hand, the total gas fractions in the progenitors of massive cluster galaxies are consistently lower since $\sim1$ Gyr after the Big Bang in comparison to field massive galaxies (Figure \ref{fig:gasfrac}).

We hence hypothesize that the reduced star formation in the massive cluster galaxies at $z\sim2$ is a delayed cumulative effect of the lower gas fractions in their progenitors due to the very environment they evolve in instead of direct interactions with other cluster galaxies or the intra-cluster medium.

\label{sec:summary}

\section{Acknowledgement}

The authors would like to thank Dr. Joel Leja for insightful feedback for the paper. The authors would also like to thank the anonymous referee for their careful reading of the manuscript which has helped improved the clarity of the work. GGK acknowledges the support of the Australian Research Council through the Discovery Project DP170103470. AG acknowledges support of the Australian Research Council Centre of Excellence for All Sky Astrophysics in 3 Dimensions (ASTRO 3D), through project number CE170100013. T.N. and K. G., acknowledge support from Australian Research Council Laureate Fellowship FL180100060

This paper includes data gathered with the 6.5 meter Magellan Telescopes located at Las Campanas Observatory, Chile and W M Keck Observatory, Hawaii. The authors also wish to recognize and acknowledge the very significant cultural role and reverence that the summit of Mauna Kea has always had within the indigenous Hawaiian community. We are most fortunate to have the opportunity to conduct observations from the summit.

\software{
	Prospector \citep{Leja2017, Johnson2019},
	EAZY \citep{Brammer2008},
	FAST \citep{Kriek2009}, 
	FSPS \citep{Conroy2009}, 
	MIST \citep{Dotter2016}}

\end{document}